\pdfoutput=1
\documentclass[prstab,groupedaddress,amsmath,reprint,floatfix]{revtex4-1}
\usepackage{graphicx,hyperref,siunitx}
\DeclareMathOperator{\arsinh}{arsinh}
\DeclareMathOperator{\atan2}{atan2}

\DeclareMathOperator{\dilog}{dilog}
\newcommand{\avg}[1]{\big\langle #1 \big\rangle}
\newcommand{\totd}{\,\mathrm d}

\begin{document}

\title{Low emittance lattice design from first principles:\\reverse bending and longitudinal gradient bends}

\author{B. Riemann}
\email{bernard.riemann@psi.ch}
\author{A. Streun}
\email{andreas.streun@psi.ch}
\affiliation{Paul Scherrer Institut, CH-5232 Villigen PSI, Switzerland}

\date{\today}

\begin{abstract}
  The well-known relaxed theoretical minimum emittance (TME) cell is commonly used in the design of multi-bend achromat (MBA) lattices for the new generation of diffraction limited storage rings. But significantly lower emittance at moderate focusing properties can be achieved by combining longitudinal gradient bends (LGB) and reverse bends (RB) in a periodic lattice unit cell. LGBs alone, however, are of rather limited gain.
  
  We investigate the emittance achievable for different unit cell classes as a function of the cell phase advance in a most general framework, i.e. with a minimum of assumptions on the particular cell optics. Each case is illustrated with a practical example of a realistic lattice cell, eventually leading to the LGB/RB unit cell of the baseline lattice for the upgrade of the Swiss Light Source.
\end{abstract}

\maketitle

\section{Introduction\label{sec:intro}}

The quantum nature of light is the origin of finite emittance in an electron storage ring: sudden loss of energy due to photon emission causes an electron to start a betatron oscillation around the closed orbit corresponding to its reduced energy. The orbit position as a function of energy is given by the lattice dispersion. Thus emittance is minimized by suppression of dispersion at locations where radiation is emitted, i.e. in the bending magnets (bends). In a planar, separate-function lattice, this can be done in three ways:
\begin{enumerate}
\item Horizontal focusing of the beam into the bends, since dispersion occurs in the horizontal dimension in a planar lattice.
\item Using many bends of small deflection angle in order to limit the dispersion growth inside the bend. This leads to the concept of the multi-bend achromat (MBA) lattice \cite{einfeld-plesko}.
\item Variation of the magnetic field inside the bend to compensate the growth of dispersion beyond the magnet center. This is the concept of the longitudinal gradient bend (LGB) \cite{wru1992}.
\end{enumerate}
For the new generation of diffraction limited storage rings, technological progress enabled miniaturization of vacuum chambers and magnets. This leads to a reduction of unit cell length, so that the double or triple bend achromats of third generation light sources could be replaced by MBAs containing five or more lattice cells within the same arc length, as pioneered by MAX~IV \cite{max4-prab}. Since the emittance $\varepsilon$ scales with the inverse cube of unit cell bending angle \cite{wiedemann,leemann-streun}, the introduction of small-aperture MBA lattices enabled emittance reduction by 1--2 orders of magnitude compared to third generation light sources.

The emittance of an MBA is dominated by the emittance of the unit cell. The two dispersion suppressor cells at the ends of the arc are similar to half unit cells. The unit cell is made from a bend and focusing elements (i.e. quadrupoles) to provide periodic solutions for beta functions and dispersion in order to string together several cells.

The requirement of a horizontal focus in the bend center for minimizing dispersion leads to a high horizontal betatron phase advance $2\phi$ of a low emittance cell. We consider phase advances $2\phi < \pi$ as sensible, because solutions with $2\phi > \pi$ require a second focus of the horizontal beta function, resulting in rather long cells.

The so-called Theoretical Minimum Emittance (TME) cell provides the minimum emittance for a unit cell containing one homogeneous bend, as can be shown without any assumptions on the particular cell optics \cite{teng-tme}, but the corresponding phase advance is very high. Therefore most lattices are based on relaxed TME cells at lower phase advance providing about three times larger emittance than the TME.

In order to further reduce the emittance, LGBs and reverse bends (RB, also called anti-bends) gained interest: LGBs have the potential to achieve sub-TME emittance by concentrating the quantum excitation in regions of vanishing dispersion, i.e. at the bend center \cite{gradientbend,clic-trapz,nagaoka-wrulich}. And (weak) RBs at the cell ends are useful to reduce the dispersion at the LGB center by manipulating the periodic solution of the dispersion function~\cite{steffen,delahaye,antibend}.

\vspace{1em}

In this work, we generalize the TME cell and the class of relaxed TME cells towards two different bends per cell with arbitrary longitudinal gradient but no transverse gradient, still not posing any assumptions on the detailed focusing in the cell. This allows a general study of the alternative concept of the RB cell with and without LGB in a common framework.

It is shown that, while RB cells and LGB cells both may have slight advantages over using a relaxed TME cell, only a combination of both (LGB/RB cell) enables superior emittance reduction and is also compatible with the requirements of MBAs built from periodic unit cells.

In the following, we first discuss unit cells with one bending magnet (sec.~\ref{sec:IsoMag} -- \ref{sec:lgb}) and then generalize to cells with two different bending magnets (sec.~\ref{sec:RbCells}). General treatment of emittance, optimal parameters and cell classes is done with a minimum of assumptions on the particular structure of the unit cell.

Design of a real cell, however, has to provide horizontal and vertical stability, has to take into account technical limitations and will strive for a minimum cell length. Therefore, the general treatment of each cell class is accompanied by the design \cite{opa} of a realistic example cell, which eventually cumulates in the present baseline design for the upgrade of the Swiss Light Source, SLS~2.0 \cite{sls2-jsr,sls2-cdr,sls2-ipac18} (sec.~\ref{sec:realcells}).

\section{Unit cells with one bend\label{sec:IsoMag}}
\begin{figure}[tbp]
  \includegraphics{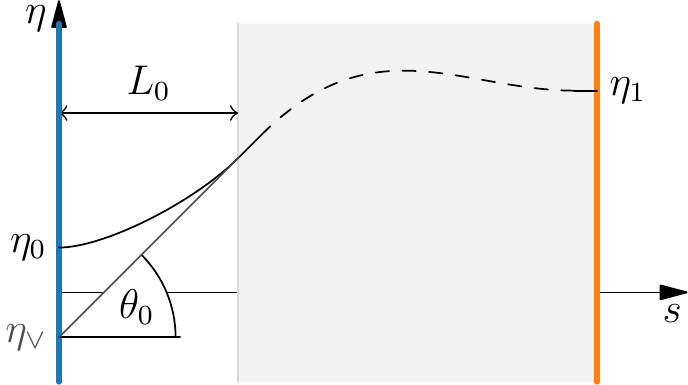}
  \caption{Course of dispersion and equivalent thin-dipole dispersion for isomagnetic half-cells. The half-cell ends are denoted by blue and orange lines. In the bend-free region (shaded area), $\beta(s)$ and $\eta(s)$ are unspecified.\label{fig:isomag_disp}}
\end{figure}

We consider a half cell of unspecified length and focusing properties, the ends of which are denoted by indices $q \in \lbrace 0, 1 \rbrace$. The half-cell ends are symmetry planes of optical functions (see Fig.~\ref{fig:isomag_disp}).

The bending magnet at position $0$ with full length $2L_0$ generates a total
bending angle $2\theta_0>0$, which for cells with only one bend equals the bending angle per cell. As this angle is small ($\theta_0 \ll 1$), the effective focal length of the bending magnet is $f = L_0 / \theta_0^2 \gg L_0$ such that $\beta(s)$ approximately propagates like in a drift space, $\beta(s) = \beta_0 + s^2 / \beta_0$ -- this is consistent with \cite{trbojevic-courant,leemann-streun,gradientbend} and confines our model to magnets without transverse gradients. Thus the phase advance in the bending magnet is always defined by $\beta_0$, resulting in a lower limit for the half-cell phase advance
\begin{align}
  \label{eq:phiLim}
  \phi &>  \arctan(L_0 / \beta_0) &\text{or}\quad \beta_0 &> L_0 \cot \phi.
\end{align}

Unlike for $\beta(s)$, the dispersion function $\eta(s)$ inside the bend depends specifically on the shape of its magnetic field, respectively the closed-orbit curvature $b(s)$, via \cite{gradientbend}
\begin{align}
  \label{eq:etadef}
  \eta(s) &= \eta_0 + \int\limits_0^s \eta^\prime(\tilde s) \totd \tilde s, & \text{with } \eta^\prime(s) &= \int\limits_0^s b(\tilde s) \totd \tilde s.
\end{align}
The dispersion outside of the bending magnet depends only on its length and bending angle, as at its ends $\eta(L_0)=\eta(-L_0)$ and  $\eta^\prime(\pm L_0) = \pm \theta_0$. It is therefore reasonable for matching purposes to introduce the equivalent dispersion value $\eta_\vee = \eta(L_0) - L_0\cdot\eta^\prime(L_0)$ at the bend center (position 0) that would occur if the bending magnet was thin, but retained its bending angle (Fig.~\ref{fig:isomag_disp}). Using partial integration, one obtains
\begin{align}
  \label{eq:etaVee} 
  \eta_\vee = \eta_0 - \int\limits_0^{L_0} s \cdot b(s) \totd s.
\end{align}

\begin{figure}[tbp]
  \includegraphics{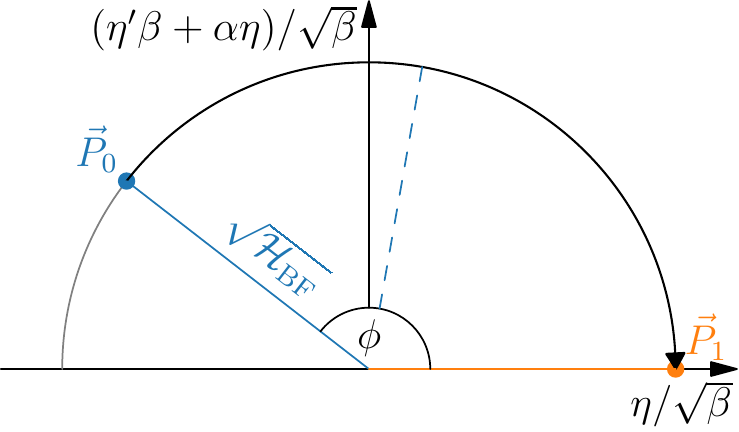}
  \caption{Normalized phase space for dispersion. For relaxed TME cells, $\vec P_0$ is shifted so that $\eta_\vee > 0$ to enforce $\phi < \pi/2$ (dashed line).\label{fig:isomag_phasecircle}}
\end{figure}

The horizontal transfer matrix of the half-cell \cite{wille} can be expressed as
\begin{align}
  \label{eq:T_half}
  \mathbf T &= \mathbf B_1 \mathbf R(\phi) \mathbf B_0^{-1} & \text{ with } \mathbf B &= \frac 1{\sqrt{\beta}}
  \begin{pmatrix}
    \beta & 0 \\ -\alpha & 1
  \end{pmatrix},
\end{align}
where $\mathbf B_q$ is the mapping from normalized phase space to standard phase space at the respective end $q$ of the half-cell, and $\mathbf R(\phi)$ is a clock-wise rotation matrix with the half-cell phase advance $\phi$. The matching condition for the thin-dipole dispersion can then be written as
\begin{align}
  \label{eq:dispmatch}
  \mathbf B_1^{-1}
  \begin{pmatrix}
    \eta_1 \\ 0
  \end{pmatrix} = \mathbf R(\phi) \mathbf B_0^{-1}
  \begin{pmatrix}
    \eta_\vee \\ \theta_0 
  \end{pmatrix}.
\end{align}
Insertion of optics expressions for the $\mathbf B_q$ matrices using symmetry conditions ($\alpha_q=0$) yields
\begin{gather}
  \vec P_1 = \mathbf R(\phi)\;\vec P_0, \quad \text{with } \nonumber \\
  \vec P_0 = \frac 1{\sqrt{\beta_0}}
  \begin{pmatrix}
    \eta_\vee  \\ \theta_0 \beta_0 
  \end{pmatrix}, \quad \vec P_1 = 
  \begin{pmatrix}
    \eta_1 / \sqrt{\beta_1} \\ 0 
  \end{pmatrix}.
  \label{eq:circle_cond}
\end{gather}
This is shown in Fig.~\ref{fig:isomag_phasecircle} with the half-cell phase advance
\begin{align}
  \label{eq:tme_phase}
  \phi &= \atan2(\theta_0 \beta_0, \eta_\vee),
\end{align}
where $\atan2$ is the four-quadrant inverse tangent, returning the signed angle of a point $(x,y)$ in the Euclidean plane with the $x$ axis. It follows that
\begin{align}
  \label{eq:tme_etavee}
  \eta_\vee = \theta_0 \beta_0 \cot \phi.
\end{align}

The previously mentioned drift-space assumption for the bend implies the horizontal damping partition $J_\mathrm x \approx 1$ \cite{wiedemann}, so that the emittance contribution of a cell is proportional to the fraction of radiation integrals \cite{sands}
\begin{align}
  \label{eq:genemit}
  \varepsilon \propto \frac{I_5}{I_2} \quad \text{with } I_5 = \int |b^3(s)| \mathcal H(s) \totd s, \\ I_2 = \int b^2(s) \totd s, \text{ and }\mathcal H(s) = \gamma \eta^2 + 2 \alpha \eta \eta' + \beta \eta^{\prime 2}  
\end{align}
being the dispersion invariant. Throughout the cell, the invariant of thin-dipole dispersion
\begin{align}
  \label{eq:H_bfr}
  \mathcal H_\mathrm{BF} = \left| \vec P_0 \right|^2 = \frac{\eta_\vee^2}{\beta_0} + \theta_0^2 \beta_0 = \left| \vec P_1 \right|^2
\end{align}
is constant (Fig.~\ref{fig:isomag_phasecircle}), and thus $\mathcal H_\mathrm{BF}$ is also the dispersion invariant in the bend-free region.

\section{(Relaxed) TME cells}

For a homogeneous bend with $b=\theta_0/L_0$, the fraction of radiation integrals from Eq.~\eqref{eq:genemit} simplifies to 
\begin{equation}
  \label{eq:tmelike}
  \frac{I_5}{I_2} = \frac{\theta_0}{L_0} \avg{\mathcal H}_0 = \theta_0^3 \avg{\hat{\mathcal H}}_0,
\end{equation}
where $\avg \cdot_0$ denotes the average over the length of the bending magnet \cite{sands}, and the normalized average of the dispersion invariant is defined via
\begin{align}
  \label{eq:normH}
  \avg{\hat{\mathcal H}}_0 = \frac{\avg{\mathcal H}_0}{L_0 \theta_0^2}.
\end{align}
Again using the drift-space approximation for $\beta(s)$ in the bending magnet and Eq.~\eqref{eq:etadef}, this average can be rewritten as (appendix \ref{app:hom}) 
\begin{align}
  \label{eq:meanH_ctme}
  \avg{\hat{\mathcal H}}_0
  =  \frac{L_0}{\beta_0} \left[ \left( \frac{\eta_0}{\theta_0 L_0} \right)^2 - \frac 13 \left( \frac{\eta_0}{\theta_0 L_0} \right) + \frac 1{20} \right] + \frac 13 \frac{\beta_0}{L_0}.
\end{align}

\subsection{The TME condition\label{sec:tme}}

Minimization of $\avg{\hat{\mathcal H}}_0$ with regard to $\beta_0,\eta_0$ yields the theoretical minimum emittance (TME) conditions \cite{teng-tme}
\begin{align}
  \label{eq:tme_beteta}
  \frac{\beta_0^\mathrm{TME}}{L_0} &= \frac 1{\sqrt{15}} \approx 0.258, & \frac{\eta_0^\mathrm{TME}}{\theta_0 L_0} &= \frac 16,
\end{align}
and thus by Eq.~\eqref{eq:etaVee} with the integral term simplifying to $\theta_0 L_0 / 2$ (see \cite{leemann-streun}), using Eqs.~\eqref{eq:tme_phase} and \eqref{eq:tmelike}
\begin{align}
  ( I_5 / I_2)_\mathrm{TME} &= \theta_0^3 \frac 2{3\sqrt{15}}, \nonumber \\
  \phi_\mathrm{TME} &= \pi - \arctan \sqrt{\frac 35} \approx 142.2^\circ.
\end{align}
The TME cell provides the minimum possible emittance for a single homogeneous bend per cell, but requires a large phase advance $2\phi$. Also, considerable focusing into the bending magnet is required to reach the necessary $\beta_0$.

For the following parts of this work, all emittances are compared to that of the ideal TME cell (implying a homogeneous bending magnet). We therefore define the emittance ratio in accordance with
\cite{antibend} as
\begin{align}
  F = \frac{I_5 / I_2}{(I_5 / I_2)_\mathrm{TME}}.
\end{align}

\begin{figure*}[t]
  \includegraphics[scale=0.85]{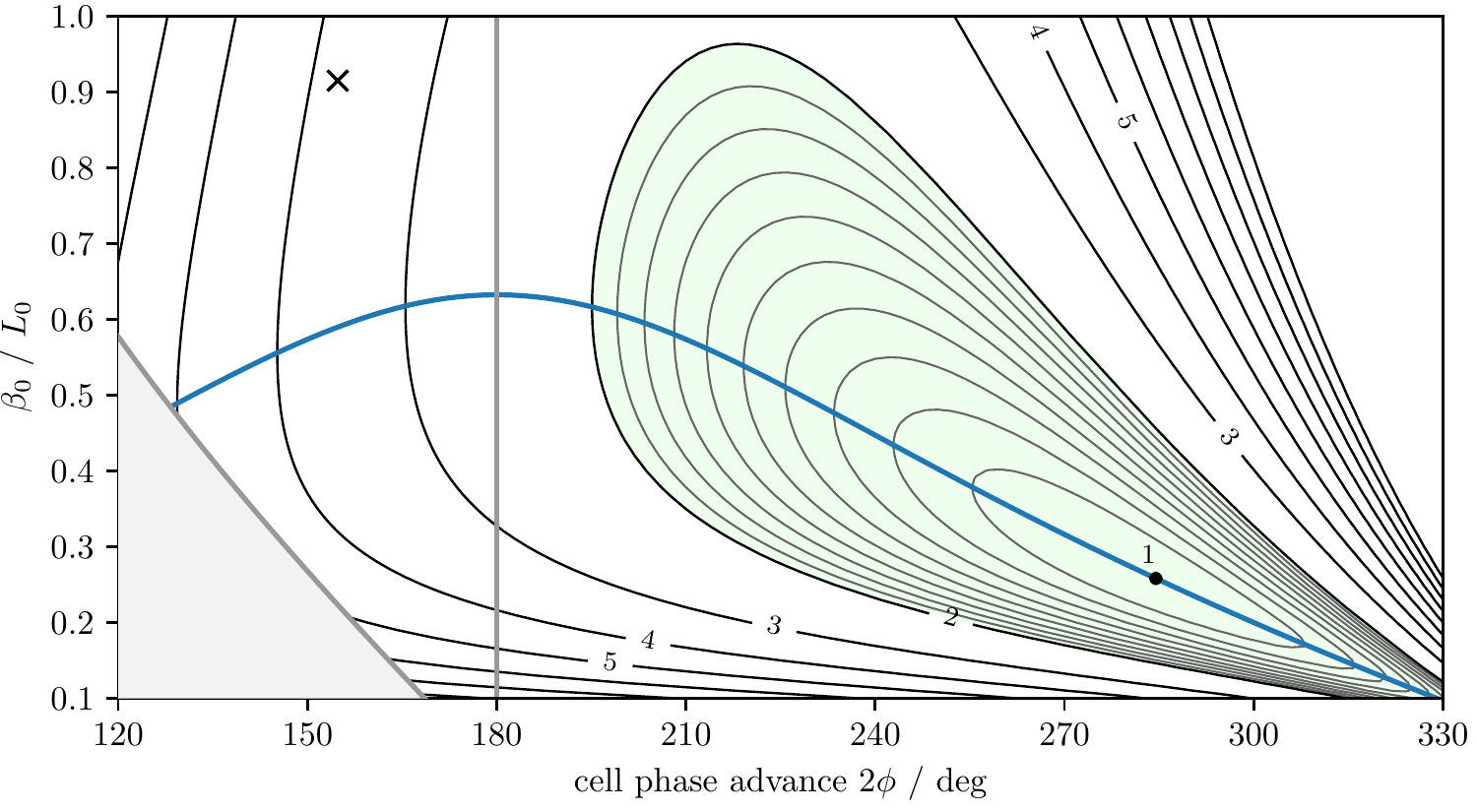}
  \caption{\label{fig:IsoMagTme}Emittance ratio $F$ in the $(\beta_0,\phi)$ plane for (relaxed) TME cells. Black iso-lines show values of $F$ in integer steps up to 10. The gray iso-lines in the green area show values $F < 2$ in steps of 0.1. The blue line shows $\beta_0^\mathrm{opt}(\phi)$, resulting in minimal emittance at a given phase. The TME condition $F=1$ is indicated by a black dot. The example cell parameters are indicated by the cross marker. Parameters in the gray area (bottom left) are not attainable.}
\end{figure*}

\subsection{Emittance in the ($\mathbf{\phi, \beta_0}$) plane\label{sec:ctme}}

To obtain minimal emittances for a relaxed TME cell depending on a fixed phase advance, one can insert the homogeneous case $b(s) = \theta_0 / L$ into Eq.~\eqref{eq:etaVee} and the phase relation following from Eq.~\eqref{eq:tme_etavee}, so that
\begin{align}
  \eta_0(\beta_0, \phi) = \frac 12 \theta_0 L_0 + \theta_0 \beta_0 \cot \phi.
\end{align}
With this expression, Eq.~\eqref{eq:meanH_ctme} can be transformed to depend on $\beta_0$ and $\phi$,
\begin{gather}
  \avg{\hat{\mathcal H}}_0(\beta_0, \phi) = \frac 2{15} \frac{L_0}{\beta_0} + A(\phi)
  \frac{\beta_0}{L_0} + \frac 23 \cot \phi\\
  \nonumber
   \text{with} \quad A(\phi) = \frac 13 + \cot^2 \phi.
\end{gather}
The optics settings for minimal emittance at a given phase then follow via
\begin{align}
  \label{eq:tmec-optphi} 
  \frac{\beta_0^\mathrm{opt}(\phi)}{L_0} &= \sqrt{\frac{2/15}{A(\phi)}}.
\end{align}

The aforementioned relations for relaxed TME cells are shown in Fig.~\ref{fig:IsoMagTme}. It is apparent that sensible phase advances $2\phi < \pi$ can only be realised at significantly higher emittances $F > 2.45$ relative to a cell fulfilling the TME condition.

\subsection{Relaxed TME example cell\label{sec:tmeExample}}

\begin{figure}[bp]
  \includegraphics[scale=0.85]{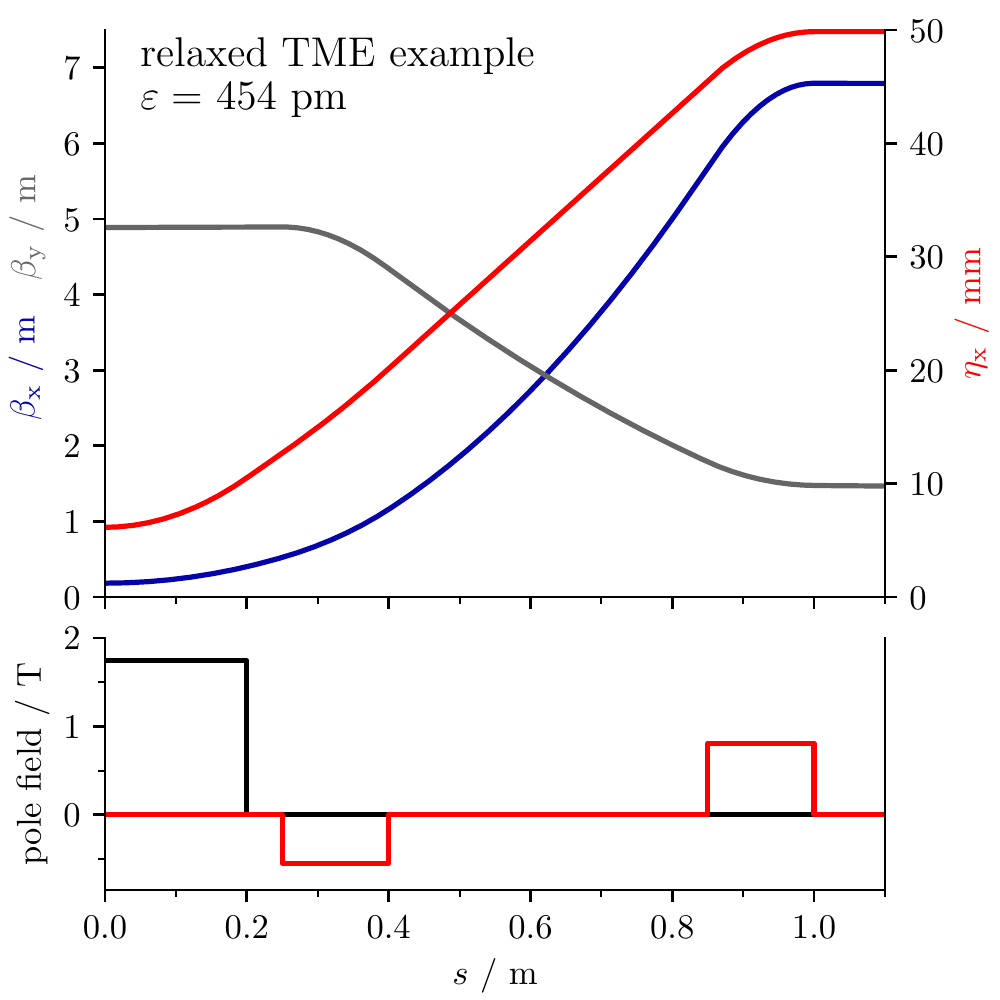}
  \caption{\label{fig:cella}Half cell of the relaxed TME type. Shown are the optical functions $\beta_\mathrm x$ (blue), $\beta_\mathrm y$ (gray) and dispersion $\eta_\mathrm x$ (red). The lower plot shows the pole-tip field components for $R=13$~mm half gap (or bore radius), dipole ($B$, black) and quadrupole ($B'R$, red).}
\end{figure}

In a series of examples throughout this work, we demonstrate the development of a lattice cell \cite{opa}, starting with a relaxed TME cell. The bend half-length was chosen as $L_0=\SI{0.2}m$, and assuming a beam energy of \SI{2.4}{GeV}, the half angle was set to $\theta_0=2.5^\circ$. For these values, the emittance of a TME cell is $\varepsilon_\mathrm{TME}=\SI{121}{pm}$.

A common technique to reduce the TME cell phase advance consists in detuning
$\eta_\vee$ to positive values so that $\phi < \pi/2$ (see
Fig.~\ref{fig:isomag_phasecircle}). The half-cell phase advances of the example cell were fixed at $\phi = \num{0.43} \pi$ in the horizontal plane and $\phi_\mathrm y = \num{0.13}\pi$ in the vertical plane -- the higher horizontal than vertical tune results from focusing into the bend in order to achieve small emittance. Fixing the half-cell length to \SI{1.1}m constrains $\beta_0$ to a larger-than-optimal value of \SI{0.183}m. The example cell parameters are also marked in the emittance surface in Fig.~\ref{fig:IsoMagTme}.

Fig.~\ref{fig:cella} shows the optical functions and the magnetic field. The emittance of this relaxed TME cell is $\varepsilon = \SI{454}{pm}$ (or $F = 3.75$).

\section{Longitudinal gradient bend cells\label{sec:lgb}}

We continue the study by replacing the homogeneous bend with an LGB. As variation of field in a longitudinal gradient bend can be chosen arbitrarily, a general closed-form solution without detailed specification of the field is at least cumbersome, and a variety of different magnet profiles for LGBs have been considered, e.g. \cite{gradientbend,clic-trapz, nagaoka-wrulich}. 
In this work, we show elementary properties of LGBs using a simple curvature function with only one free parameter (sec.~\ref{sec:idm}), and then generalize some properties to numerically optimized free-form LGBs (sec.~\ref{sec:freePlus}).

Our description of the emittance contribution from longitudinal gradient bends closely follows \cite{gradientbend}. To describe the variation of curvature in the bend, we define a normalized curvature function as
\begin{align}
  \label{eq:hatb}
  \hat b(s) &= \frac{L_0}{\theta_0} b(s), & \text{ with } \avg{\hat b(s)} &= 1. 
\end{align}
so that the emittance integrals in the bends can be expressed using averages via
\begin{align}
  \label{eq:I520}
  I_5^{(0)} &= \frac{|\theta_0|^5}{L_0} \avg{|\hat b^3| \hat{\mathcal H}}_0, & 
  I_2^{(0)} &= \frac{\theta_0^2}{L_0} \avg{\hat b^2}_0,
\end{align}
and Eq.~\eqref{eq:genemit} simplifies to
\begin{align}
  \label{eq:f_lgb}
  \frac{I_5}{I_2} &= \theta_0^3 \frac{\avg{|\hat b^3| \hat{\mathcal H}}_0}{\avg{\hat b^2}_0}. 
\end{align}
The numerator expression can be written as (appendix \ref{app:eint})
\begin{align}
  \avg{|\hat b^3| \hat{\mathcal H}}_0 &=  C \cdot \frac{\beta_0}{L_0} + \frac{L_0}{\beta_0} \cdot D\left(\frac{\eta_0}{\theta_0 L_0}\right) \label{eq:lgb_b3H} \\
  \text{with} \quad C &= \avg{|\hat b^3| \left( \frac{\eta'}{\theta} \right)^2}, \nonumber \\
  D(x) &= \avg{|\hat b^3|}\;x^2 - 2 \avg{|\hat b^3|v}\;x + \avg{|\hat b^3|v^2}. \nonumber
\end{align}
After division by the denominator term $\avg{\hat b^2}$, the magnet-specific coefficients of the $D(x)$ polynomial (appendix \ref{app:eint}) and the coefficient $C$ replace constant values in the description of the homogeneous bend in Eq.~\eqref{eq:meanH_ctme}. These parameters are dimensionless variants of the $\mathcal I_n$ terms for the symmetric bend in \cite{gradientbend}.

Based on Eq.~\eqref{eq:etaVee}, a further magnet-specific parameter $V$ is required to normalize the difference between the dispersion value $\eta_0$ at the center bend and the equivalent thin-dipole dispersion $\eta_\vee$ \cite{leemann-streun} 
\begin{align}
  \label{eq:lgb_etavee}
  \eta_\vee &= \eta_0 - V\;\theta_0 L_0 & \text{with } V = \avg{\hat b \frac s{L_0}}.
\end{align}
To characterize the concentration of magnetic field
respectively curvature in the central bend region, we introduce the field
enhancement factor \cite{gradientbend}
\begin{align}
  R &= \frac{\max b(s)}{\avg{b(s)}} = \hat b(0).
\end{align}
The four quantities $C, D, V, R$ fully describe the radiation and optics properties of the LGB in our model. The calculation of all required magnet-specific variables from the normalized
curvature $\hat b(s)$ is shown in appendix \ref{app:eint}.

\subsection{Emittance in the ($\mathbf{\phi, \beta_0}$) plane}

To obtain the emittance for a given phase advance of the cell,
one inserts the phase relation following from Eqs.~\eqref{eq:lgb_etavee} and \eqref{eq:tme_etavee},
\begin{align}
  \label{eq:et0phi}
  \eta_0(\beta_0,\phi) = V \theta_0 L_0 + \theta_0 \beta_0 \cot \phi, 
\end{align}
into Eq.~\eqref{eq:lgb_b3H} so that
\begin{align}
  \label{eq:lgb_b3Hphi}
  \avg{|\hat b^3| \hat{\mathcal H}}(\beta_0,\phi) &= \frac{L_0}{\beta_0} D(V) + A(\phi) \frac{\beta_0}{L_0} + \tilde A \cot \phi\\
  \nonumber \text{with} \quad A(\phi) &= C + \avg{|\hat b^3|} \cot^2 \phi, \\
  \nonumber \tilde A &= 2 \left( \avg{|\hat b^3|} V - \avg{|\hat b^3|v}  \right).
\end{align}

The calculation of optimal $\beta_0(\phi)$ is analogous to Eq.~\eqref{eq:tmec-optphi}. Substituting this result into Eq.~\eqref{eq:lgb_b3Hphi} yields
\begin{align}
  \avg{|\hat b^3| \hat{\mathcal H}}^\mathrm{opt}(\phi) &= 2 \sqrt{D(V) A(\phi)} + \tilde A \cot \phi.
\end{align}
The optimal phase can be derived from allowing the derivative towards $\cot \phi$ to vanish, so that
\begin{align}
  2 \avg{|\hat b^3|} \sqrt{D(V)} \cot \phi_\mathrm{opt} = -\tilde A \sqrt{A(\phi_\mathrm{opt})}.
\end{align}
It can be shown (appendix \ref{app:eint}) that for $b(s) \geq 0$, $\tilde A >
0$. As all other quantities but $\cot \phi$ are also positive, this implies in
our context that the optimal cell phase advance $2\phi$ is larger than $\pi$ for
any LGB cell with only positive curvature in its bend. In this case,
\begin{align}
  \label{eq:lgb_phiOpt}
  \phi_\mathrm{opt} = \pi - \arctan \sqrt{\frac{4 D(V) \avg{|\hat b^3|}^2}{C \tilde A^2} - \frac{\avg{|\hat b^3|}}{C}}.
\end{align}

\subsection{An elementary LGB magnet (IDM)\label{sec:idm}}

\begin{figure*}[t]
  \includegraphics[scale=0.85]{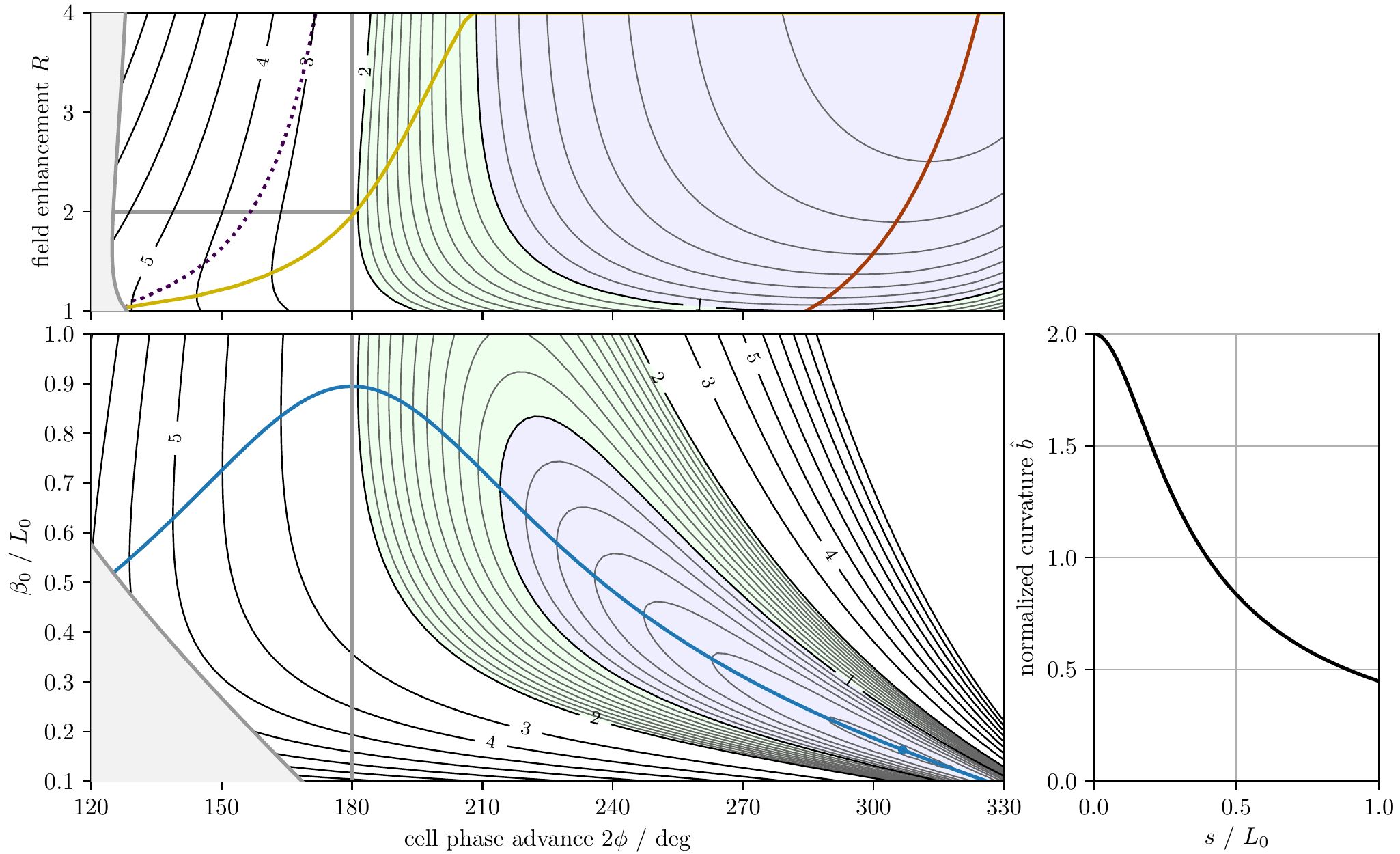}
  \caption{Properties of the IDM shape. Top: TME emittance ratio $F$ for optimal $\beta_0$ in dependence of $R,\phi$. The red line shows $\phi_\mathrm{opt}$ for given $R$. The dark-yellow line shows the optimal field enhancement $R$ for given $\phi$.  Above the dotted line (critical field enhancement), a relaxed TME cell tuned to the LGB-optimal parameters $\phi,\beta_0^\mathrm{opt}$ yields lower emittance than the LGB cell. The region with $F < 1$ is shaded in blue (see legend in Fig.~\ref{fig:IsoMagTme}).\label{fig:idm_emit}
Bottom Left: $F$ in the $(\beta_0,\phi)$ plane for an IDM with $R=2$; the optimum emittance is denoted with a blue dot.\label{fig:IsoMagIdm}
Right: IDM on-axis field profile for $R=2$.\label{fig:IsoMagIdmYoke}}
\end{figure*}

To proceed further, we need to specify the shape of longitudinal field variation $\hat b(s)$. We use the curvature function
\begin{align}
  \hat b(s) = \frac R{w(s)} \text{ for } s \leq L_0,
\end{align}
where $w(s)=\sqrt{1+(s/h)^2}$ is a normalized distance to a point in the magnet mid-plane $s=0$ with a transverse offset $h$. Due to the inverse dependence of curvature on this distance, the shape is named ``inverse distance-scaling magnet'' (IDM) shape in the following.

The IDM curvature has the advantage of being differentiable for all values $|s| < L_0$, leading to a smooth yoke shape (see e.g.~\cite{tanabe}). For $s>L_0$, the curvature vanishes. It thus can in principle be realized 'as-is' when not considering fringe fields at the magnet end. 

The field enhancement factor for the IDM shape is then given by
\begin{align}
  R = \frac{L_0/h}{\arsinh(L_0/h)}.
\end{align}
In the limit $h \rightarrow \infty$, equivalent to $R \rightarrow 1$, the IDM reduces to a homogeneous
magnet of length $L_0$ (sec.~\ref{sec:tme}). Emittance integrals and related quantities are computed in appendix \ref{app:eint}.

The properties of LGB cells utilizing IDMs in dependence of $R$ are shown in Fig.~\ref{fig:idm_emit}. For optimized cell phase advance $\phi$ and $\beta_0$, the emittance of the LGB cell relative to a relaxed TME cell with equal $\phi$, $\beta_0$ can be significantly reduced for increasing $R$. This optimal phase advance unfortunately increases with $R$, such that any given phase in the region of interest, there exists a critical field enhancement factor above which the LGB cell emittance is actually larger than that of a relaxed TME cell. It is disadvantageous that this critical field enhancement decreases for lower cell phase advances, which are of special interest. We can also observe that in the region of interest, $F < 2$ is not possible. On the other hand, we observe that $R \leq 2$ is sufficient. This missing emittance reduction is a result of the improperly matched optics at the bend for sensible phase advances (especially the lower dispersion bound $\eta_0 > V \theta_0 L_0$ for $2\phi < \pi$).

\begin{figure*}[t]
  \includegraphics[scale=0.85]{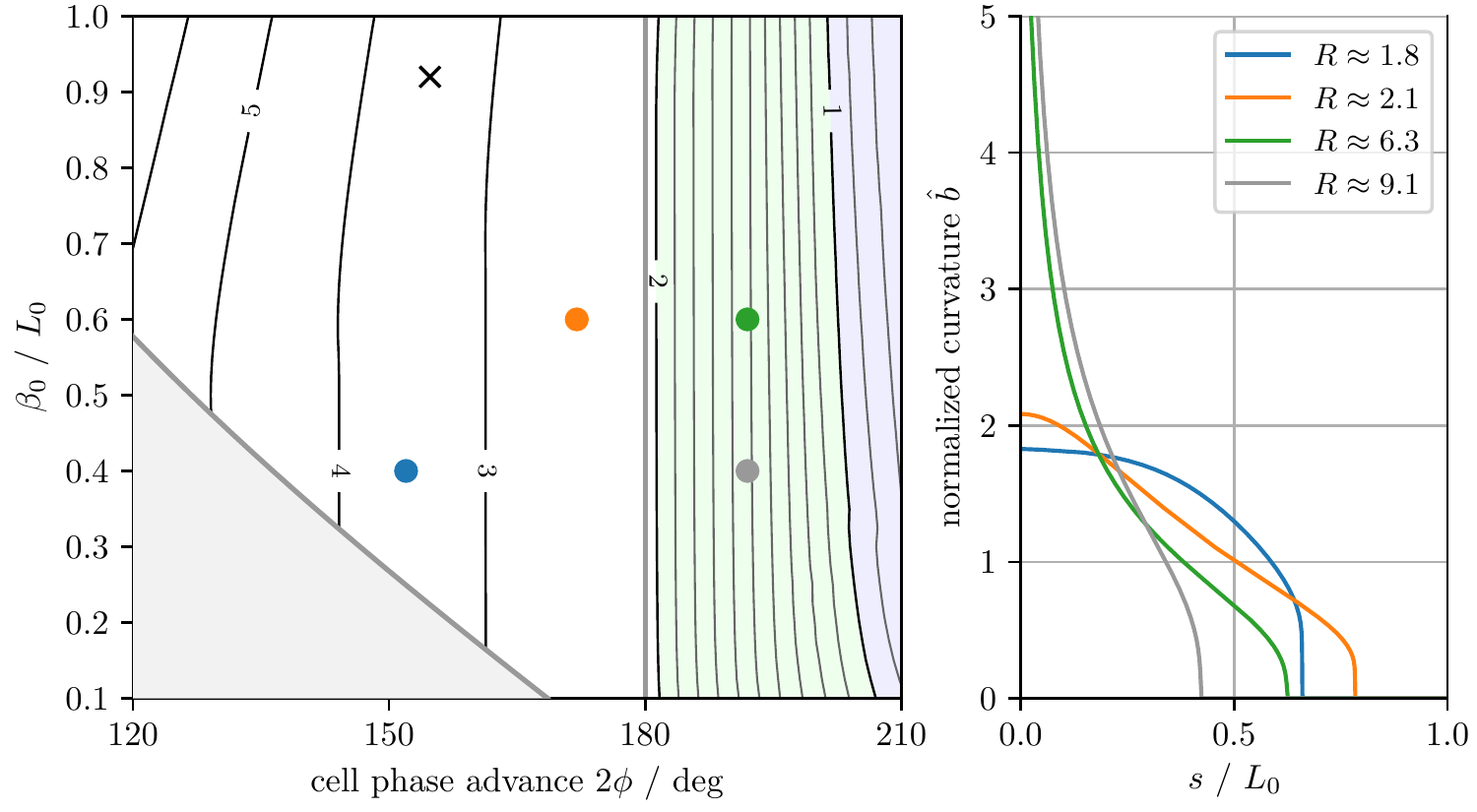}
  \caption{TME emittance ratio $F$ and example curvatures for free-form LGB cells
    with the constraint of positive curvature $b(s) \geq 0$.
    Left: $F$ in dependence of $\beta_0, \phi$ (see legend in Fig.~\ref{fig:IsoMagIdm}). Colored markers denote the curvature examples in the right plot.
    Right: curvature examples for setups marked in left plot.\label{fig:IsoMagOptBounded}}
\end{figure*}

To select a proper field enhancement for a study of the ($\phi, \beta_0$) plane, we choose $R = 2$ which is moderate and technically feasible for many setups and in principle allows significant emittance reduction. The characteristics of the LGB cell with this field enhancement factor are shown in Fig.~\ref{fig:IsoMagIdm} (bottom left). On one hand, the capability of the LGB cell in reducing emittance for large phase advances relative to relaxed TME cells is again obvious. On the other hand, the emittance reduction for sensible phase advances is marginal, although having the advantage of being robust towards increasing $\beta_0$.

\subsection{Optimized free-form LGBs with positive curvature\label{sec:freePlus}}

Is the limited performance of the IDM magnet for sensible phase advances a phenomenon that generalizes to all LGBs? For given $\phi,\beta_0$, we strive to find a curvature shape with $b(s) \geq 0$ that minimizes $F$, respectively the functional $F(b)$. This is a continuation of a numerical study in \cite{streun-ipac15}, but with the phase-matching condition (using $\eta_\vee$) allowing to use the half-cell phase $\phi$ as a fixed parameter.

There is an infinite space of possible field shapes $b(s)$, and thus we need to apply reasonable assumptions for the following search. For the field shape to be a physical solution, it should be possible to create it as a perturbation of a homogeneous bend, and it should thus be accessible by local optimization, using the parameters of a homogeneous bend as initial values. 

The optimal shape can be approximated numerically by discretising $\hat b(s)$ into values $\underline b_q$ (details in appendix \ref{app:freeform}). We apply automatic differentiation \cite{autograd, autograd-phd} to obtain the gradient of the objective function $F(\underline b_1, \dots \underline b_Q)$. The objective function and its gradient are used as input to the limited-memory BFGS optimization algorithm \cite{l-bfgs-b}.

The optimization is carried out independently for each point on a grid in the ($\phi, \beta_0$) plane. For all points with the largest value of $\beta_0/L_0=1$, the initial values are set to $\underline b_q = 1$, equivalent to a homogeneous bend. For all other points, the optimized shape from the next-larger $\beta_0/L_0$ value at equal $\phi$ is used for initialization, requiring the optimization loop along the $\beta_0$ dimension to be carried out in reverse order. To prevent the numerical discretization of $b(s)$ from influencing convergence at very high field enhancement factors, we limit the scope of our optimization study to $2\phi \leq 210^\circ$.

\begin{figure}[b]
  \includegraphics[scale=0.85]{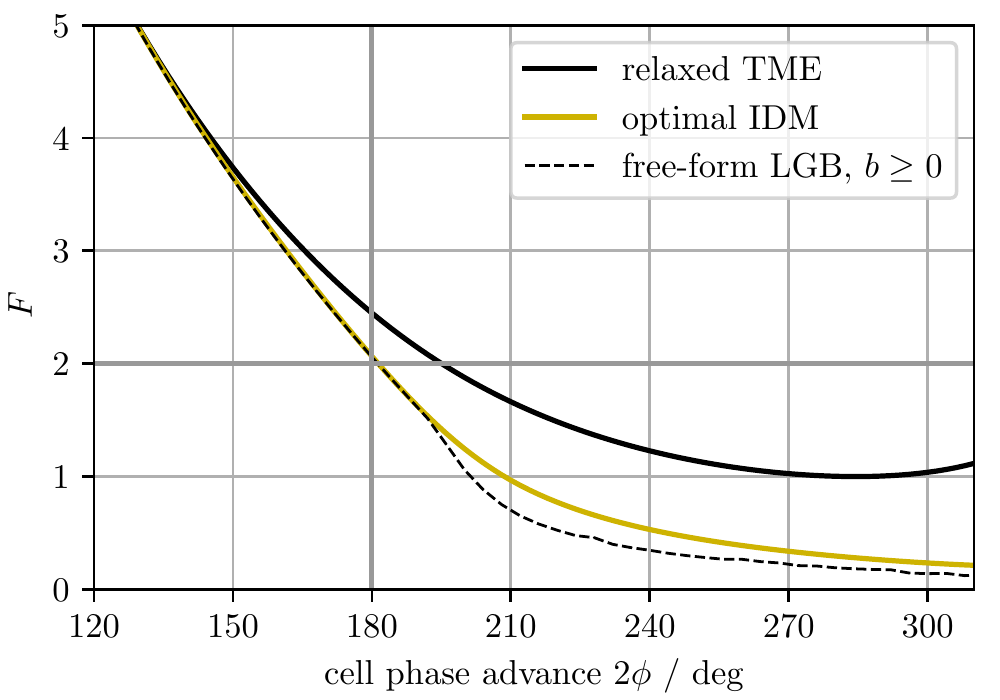}
  \caption{Comparison of minimum possible emittances for (relaxed) TME cells, free-form LGB cells with positive curvature, and LGB cells using the IDM shape.\label{fig:IsoMagLgbTmeF}}
\end{figure}

The results of this computation are shown in Figs.~\ref{fig:IsoMagOptBounded} and
\ref{fig:IsoMagLgbTmeF}.
Emittance in the ($\phi,\beta_0$) plane is always reduced relative to the TME case (Fig.~\ref{fig:IsoMagTme}), and the achieved emittance is robust regarding changes in $\beta_0$: due to the length-related quantities $\beta,\eta$ in our model scaling with $L_0$, one can always find a similar $F$ value when reducing $\beta_0$ and the $s$ dimension of $\hat b(s)$ simultaneously while keeping its integral constant
(``squeezing''). The optimization shows this effect if lower values of $F$ cannot be found by reducing $\beta_0$ and leads to reduced magnet lengths (see example shapes in Fig.~\ref{fig:IsoMagOptBounded}).

On one hand, significant emittance reduction is possible for large phase advances, where high field enhancement factors occur for the magnet shapes (like for the IDM special case). On the other hand, within our assumptions we observe that for sensible phase advances $2\phi < \pi$, no optimized shape leads to a TME emittance ratio $F \leq 2$. In this regime, IDM shapes with $R \leq 2$ at their optimal $\beta_0$ values actually yield comparable performance to the optimized LGB shapes (Fig.~\ref{fig:IsoMagLgbTmeF}).

\subsection{LGB example cell\label{sec:lgbExample}}

As a continuation of the example from sec.~\ref{sec:tmeExample}, a longitudinal gradient bend is introduced by
insertion of the appropriate optimized field profile for given $\beta_0$ and
$\phi$ as obtained in the last section. After insertion of the modified bend,
slight modifications of quadrupole strengths are required to maintain
$\phi,\phi_\mathrm y$ at their previous values (with almost equal $\beta_0=\SI{0.184}m$, example parameters marked in Fig.~\ref{fig:IsoMagOptBounded}).
\begin{figure}[b]
  \includegraphics[scale=0.85]{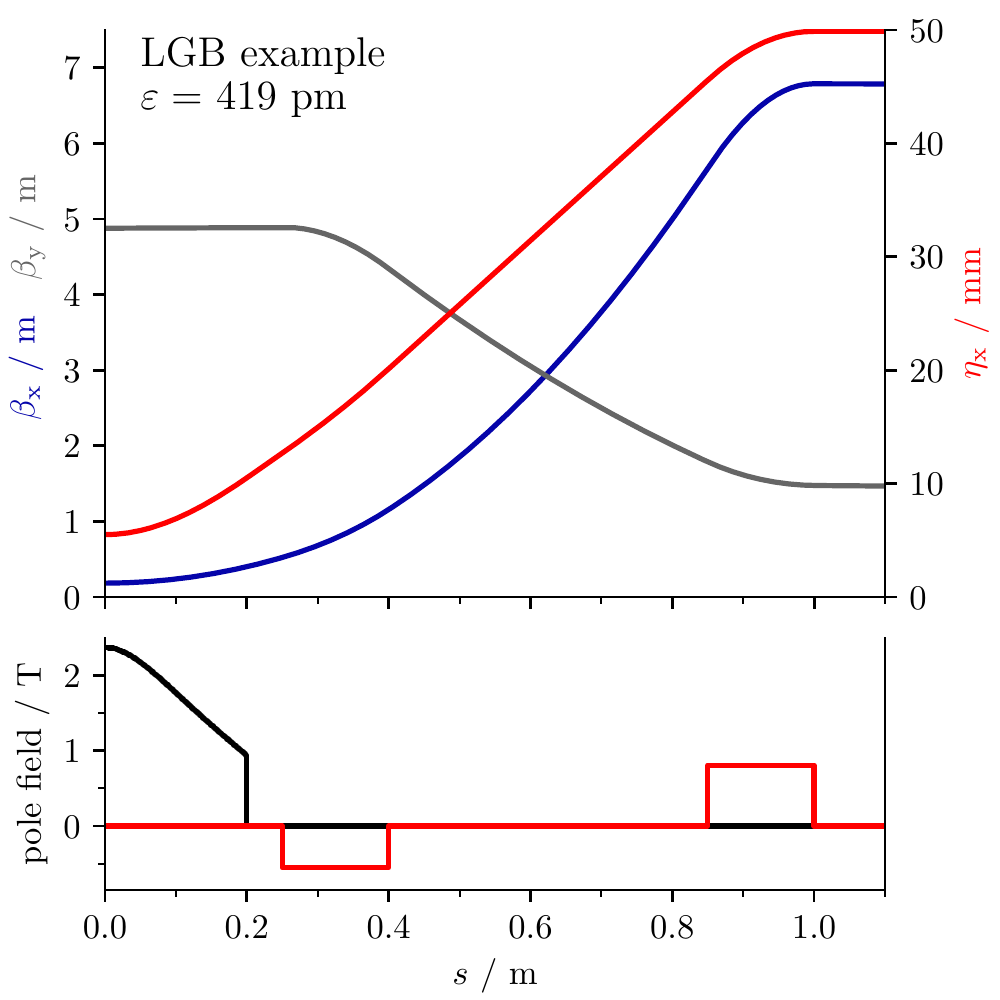}
  \caption{\label{fig:cellb}Example half cell with optimization of longitudinal field variation ($b \geq 0$) in its bend. (See Fig.~\ref{fig:cella} for legend.)}
\end{figure}
The cell optics and magnet characteristics are shown in Fig.~\ref{fig:cellb}.
The LGB field possesses a moderate enhancement factor of $R \approx 1.4$ and
leads to an emittance $\varepsilon = \SI{419}{pm}$ (or $F=\num{3.46}$) of this
example LGB cell, which is a marginal improvement by 7.7~\% relative to the
relaxed TME example (this reduction is mainly produced by an increase of $I_2$
rather than reduction of $I_5$, see Fig.~\ref{fig:histo}). Both findings are consistent with our model predictions regarding sensible phase advances $2\phi < \pi$.

\subsection{Optimized free-form LGBs with arbitrary curvature\label{sec:optUnbounded}}
\begin{figure*}[!t]
  \includegraphics[scale=0.85]{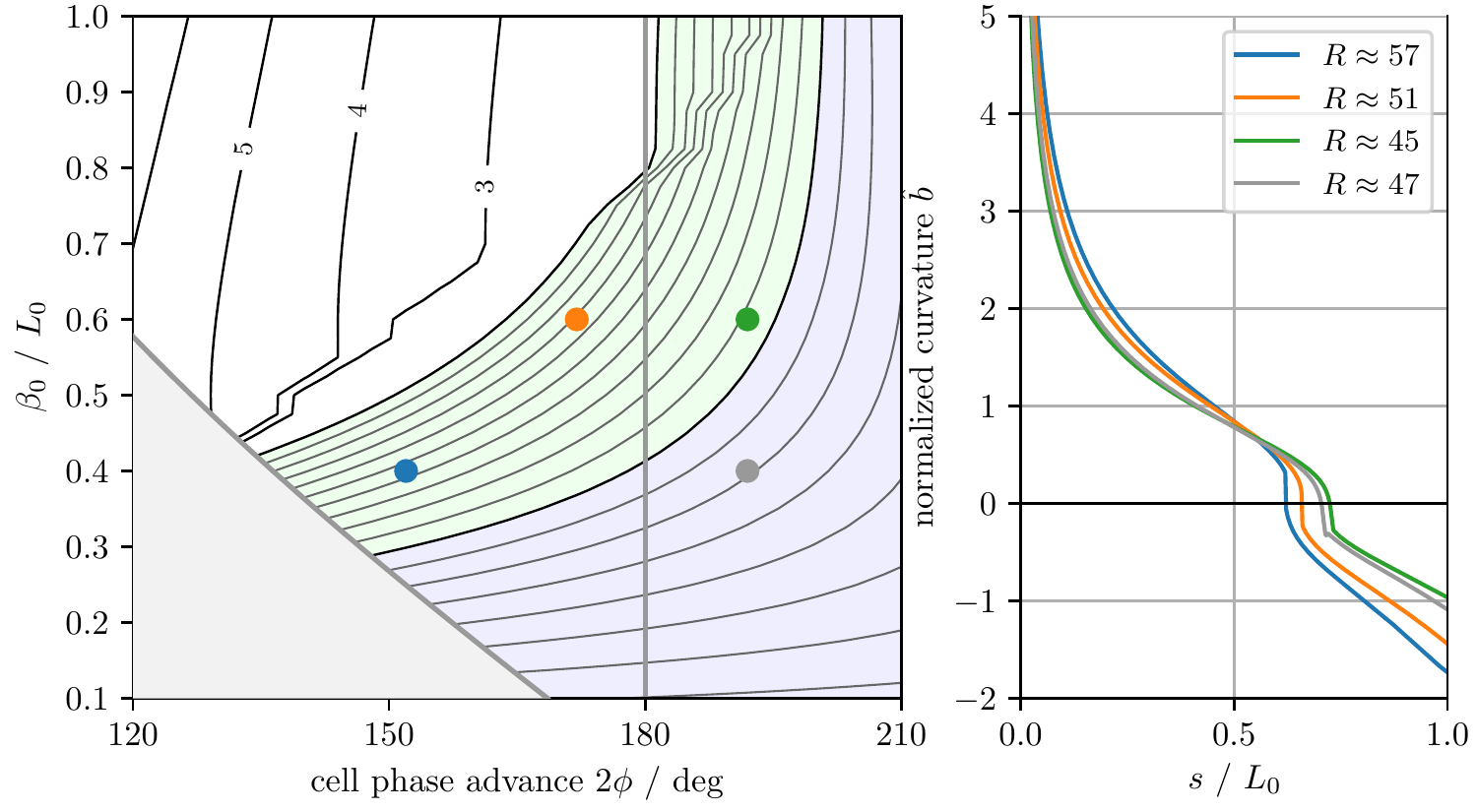}
\caption{TME emittance ratio $F$ and example curvatures for free-form LGB cells without sign constraints on $b(s)$ (see Fig.~\ref{fig:IsoMagOptBounded} for legend). \label{fig:IsoMagOptUnbounded}}
\end{figure*}

Constraining the bend to possess only positive curvatures $b(s) \geq 0$ and sensible phase advances $2 \phi < \pi$ causes $\eta_0 > V \theta_0 L_0 > 0$ to have a finite, positive lower bound. In consequence, the dispersion invariant $\mathcal H(s)$ in the bending magnet also has a lower bound -- enhancing the peak field of an LGB can only minimize $I_5$ and thus the emittance to a given extent. To investigate this limit in the following, the constraint $b(s) \geq 0$, which has been applied in the previous optimization of the free-form LGB shapes, is removed.

The full optimization procedure is again performed on the curvature shapes using L-BFGS-B \cite{l-bfgs-b}, but without application of bounds on $\underline b_q$. We obtain significantly lower emittances in the region $2 \phi < \pi$ compared to the case of positive curvatures (compare Fig.~\ref{fig:IsoMagOptUnbounded} with Fig.~\ref{fig:IsoMagOptBounded}) -- indeed emittances $F < 2$ and even $F < 1$ are possible, although seemingly large field enhancement is required.

The interesting regions in Fig.~\ref{fig:IsoMagOptUnbounded} (left), where the phase advance is sensible ($2\phi<\pi$) and small emittance ($F \sim 1$) is obtained, are adjacent to the region where all phase advance is contained inside the bend (gray shaded area defined by Eq.~\eqref{eq:phiLim}). This means that the bend basically fills the cell. 
As visible in Fig.~\ref{fig:IsoMagOptUnbounded} (right), the curvature switches sign -- one may interpret this behaviour as the free-form LGB actually splitting into a main bend of positive curvature and a {\em reverse bend} (RB) of negative curvature. There is no focusing element located between the bends, so that $\beta(s)$ behaves like in a drift space and is only refocused at the cell end. The focusing element could be imagined as a thin quadrupole as studied in \cite{antibend}: in this idealized setup, $\beta(s)$ also behaves like in a drift space over the full cell length, containing bends of opposite polarity.

The significant reduction at sensible phase advance is due to a combined action of LGB and RB. With the RB at the end of the cell, the phase advance between RB and main bend is not much less than $90^{\circ}$, so a negative kick on dispersion $-\Delta \eta_1^\prime$ applied by the RB translates in a reduction of $\eta_0$, and with it ${\cal H}_0$, at the main bend center, which enables the LGB to efficiently suppress emittance by adjusting the curvature $b(s)$ appropriately.

\begin{figure}[bt]
  \includegraphics{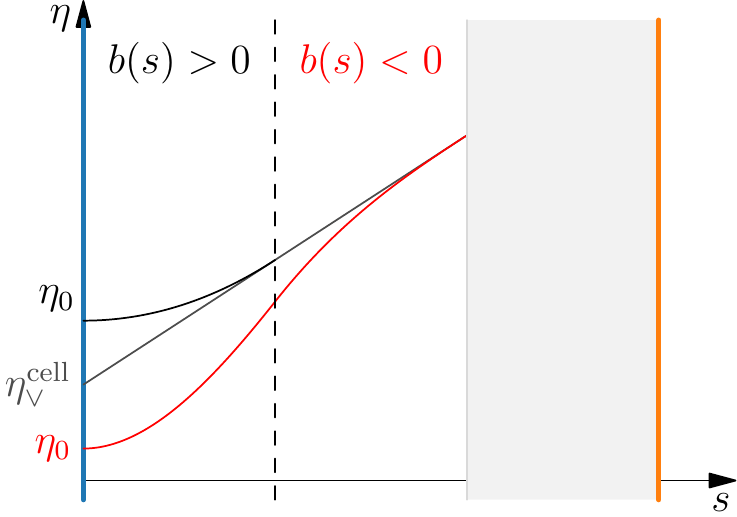}
  \caption{Behaviour of dispersion in an LGB cell for constant $\beta_0$ and $\phi < \pi / 2$ (and thus $\eta_\vee > 0$) with only positive curvature (black) and with alternating curvature (red). The reverse curvature allows $\eta_0 < \eta_\vee$. To keep the total bending angle constant including negative curvature, the overall positive curvature is increased.\label{fig:disp_elemrb}}
\end{figure}

Unboundedness of $b(s)$ means that $V$ and thus $\eta_0$ can be chosen more or less freely for a given transfer matrix by adjusting the RB acting as lever at the cell end (Fig.~\ref{fig:disp_elemrb}). We also note from the optimized shapes in Fig.~\ref{fig:IsoMagOptUnbounded} that the reverse-bend strength is largest near the magnet end, where $\beta(s) \propto s^2 $ reaches its maximum. This is reasonable as perturbations in orbits and dispersion scale with $\sqrt \beta$, making adjustments there more effective.

\section{Reverse-bend cells\label{sec:RbCells}}

Since reverse bends naturally emerge from the optimization of the free-form LGB and reveal the potential to realize $F<2$ emittance at sensible phase advance ($2\phi<\pi$), we proceed with a study of a modified unit half-cell composed from two discrete magnets at its ends, which may possess opposite polarities. Unit cells including reverse bends have been considered for a wiggler storage ring \cite{steffen}, are an established concept used for damping rings \cite{delahaye} and have recently been suggested for use in modern synchrotron light sources \cite{antibend}.

\begin{figure}[bp]
  \includegraphics{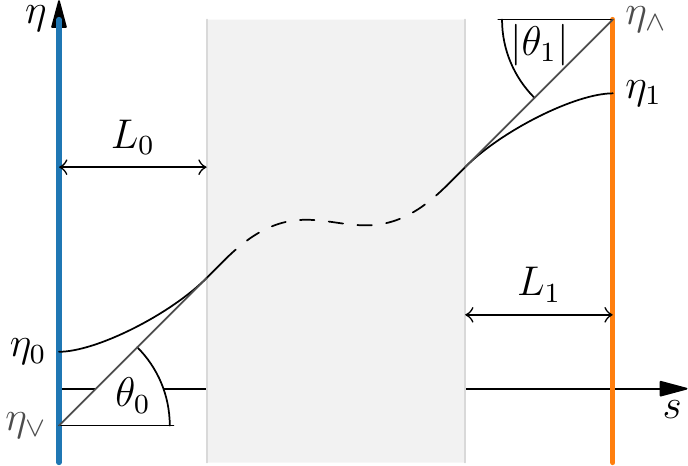}
  \caption{Dispersion $\eta_0, \eta_1$ and equivalent thin-dipole dispersion $\eta_\vee,
    \eta_\wedge$ for a half cell with unspecified interior and $\theta_1 < 0$ (compare to Fig. \ref{fig:isomag_disp}).\label{fig:twobend_disp}}
\end{figure}

A generalization of the dispersion matching condition from sec.~\ref{sec:IsoMag} to cells with two bends allows to keep the following study free from the specification of focusing elements. A second bending magnet with full length $2 L_1$ and a total bending angle $2\theta_1 \neq 0$ is introduced at the opposite half-cell end (see Fig.~\ref{fig:twobend_disp}). We define the equivalent thin-kick dispersion for the second bend in analogy to Eq.~\eqref{eq:etaVee} as
\begin{align}
  \eta_\wedge = \eta_1 - \int\limits_0^{L_1} s \cdot b(s) \totd s.
\end{align}
Then the condition for dispersion matching Eq.~\eqref{eq:dispmatch} modifies to
\begin{align}
  \mathbf B_1^{-1}
  \begin{pmatrix}
    \eta_\wedge \\
    -\theta_1
  \end{pmatrix} = \mathbf R(\phi) \mathbf B_0^{-1}
  \begin{pmatrix}
    \eta_\vee \\ \theta_0 
  \end{pmatrix},
\end{align}
so that Eq.~\eqref{eq:circle_cond} holds with a more general expression for
\begin{align}
  \label{eq:Pone_reverse}
  \vec P_1 = \frac 1{\sqrt{\beta_1}}
  \begin{pmatrix}
    \eta_\wedge \\ -\theta_1 \beta_1 
\end{pmatrix}.
\end{align}

\begin{figure}[bt]
  \includegraphics{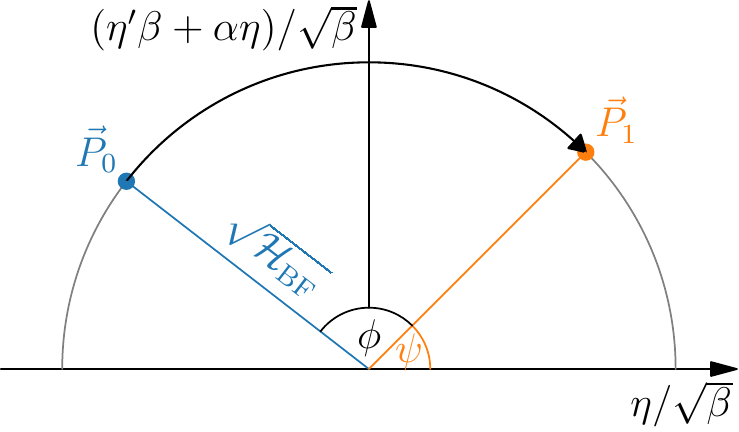}
  \caption{Normalized phase space for dispersion in a two-bend cell with $\theta_0 > 0,
    \theta_1 < 0$ (reverse-bend cell).\label{fig:twobend_phasecircle}}
\end{figure}

\noindent The dispersion matching in this general cell is sketched in Fig.~\ref{fig:twobend_phasecircle} with the
additional definition
\begin{align}
  \label{eq:rb_phase}
  \psi &= \atan2(-\theta_1 \beta_1, \eta_\wedge),
\end{align}
so that when denoting the cell phase advance from Eq.~\eqref{eq:tme_phase} as
$\phi_\mathrm{iso}$, we obtain modified phase advance and dispersion
expressions
\begin{align}
  \phi &= \phi_\mathrm{iso} - \psi, & \eta_\vee &= \theta_0 \beta_0 \cot \phi_\mathrm{iso}, \\
  \theta_1 \sqrt{\beta_1} &= -\sqrt{\mathcal H_\mathrm{BF}} \sin \psi, & 
  \eta_\wedge &= -\theta_1 \beta_1 \cot \psi.
\end{align}
The phase advance of a two-bend cell is
reduced relative to an isomagnetic cell with identical $\beta_0, \eta_\vee$ values only for a reverse-bend cell (or anti-bend cell) \cite{antibend}, as
\begin{align}
  \psi \in ]0,\pi[ \quad \Leftrightarrow \quad -\theta_1 \beta_1 > 0 \quad \Leftrightarrow \quad \theta_1 < 0
\end{align}
so that the additional dipole magnet bends the beam in opposite
direction relative to the main bend.

As transverse kicks scale with $\sqrt{\beta}$, we expect a relation between $\beta_1$ and reverse bend angle $|\theta_1|$. For larger $\beta_1$, the reverse bend will become more effective, requiring less bending angle and thus contributing less to the emittance, which naturally makes optics settings with large $\beta_1$ beneficial for reverse-bend cells. In general, however, the parameter $\beta_1$ is limited towards large values by optics constraints (e.g.~cell length, see also sec.~\ref{sec:realcells}), which need to be balanced with the available emittance reduction.

\subsection{Relative length of bending magnets}

As $\eta(s)$ and $\beta(s)$ in the RB region are rather large and do not vary much, the gain from a longitudinal field variation would be negligible, and one can therefore assume the RB to be homogeneous. In consequence, its radiation properties are fully characterized by Eq.~\eqref{eq:meanH_ctme} when replacing the index $0$ with
$1$, 
\begin{align}
  \avg{|\hat b^3| \hat{\mathcal H}}_1 &= \avg{\hat{\mathcal H}}_1 
  & \text{with} \quad \eta_1 &= \eta_\wedge + \frac 12 \theta_1 L_1.
\end{align}
Then the ratio of radiation integrals depends on the ratio of bending
magnet lengths,
\begin{align}
  \frac{I_5^{(0)} + I_5^{(1)}}{I_2^{(0)} + I_2^{(1)}} = \theta_0^3 \frac{\avg{|\hat b^3| \hat{\mathcal H}}_0 + (L_0/L_1) |\hat\theta|^5\avg{\hat{\mathcal H}}_1}{\avg{\hat b^2}_0 + (L_0/L_1) \hat\theta^2}, \\
  \text{with } \hat\theta = \theta_1 / \theta_0. \nonumber
\end{align}

The maximum orbit curvature in a cell is defined by its value at the center of the main bend, $\max b = R \theta_0 / L_0$. One can assume that this maximum achievable curvature is constant for all magnets due to technical limitations. To obtain an estimate independent of the field enhancement factor $R$, and to allow for some leverage as the large curvatures in LGB central regions are only reached using considerable design efforts, we assume the technically reasonable maximum absolute curvature in a reverse bend to be $\theta_0 / L_0$, and that the reverse bend is a homogeneous bend. In order to save space, it is therefore useful to scale the magnet lengths with their absolute bending angle as
\begin{align}
  \label{eq:L0L1}
  L_1 / L_0 = |\hat\theta|. 
\end{align}
When space is available in the unit cell, it is often beneficial to increase the reverse-bend length $L_1$ (see sec.~\ref{sec:RbExample}), as this can further reduce the emittance relative to the following calculation.

To compare emittances of unit cells with two bends with that of the TME cell, their half-cell bending angles should be equal. Defining this angle as $\Theta = \theta_0 + \theta_1$ for unit cells with two bends, we use the relation $\Theta / \theta_0 = 1 + \hat\theta$ to obtain the TME emittance ratio of a general unit cell with two bends with lengths defined by Eq.~\eqref{eq:L0L1} as
\begin{widetext}
\begin{align}
  \label{eq:fullemit}
  F &= \frac{3\sqrt{15}}2 \frac{\avg{|\hat b^3| \hat{\mathcal H}}_0 + \hat\theta^4 \avg{\hat{\mathcal H}}_1}{\left(1 + \hat\theta\right)^3 \left( \avg{\hat b^2}_0 + |\hat\theta|\right)}, \quad \text{with} \\
  \hat \theta^4 \avg{\hat{\mathcal H}}_1 &= \frac{L_0}{\beta_1}
  \left[ |\hat\theta| \left( \frac{\eta_1}{\theta_0 L_0} \right)^2 - \frac{\hat\theta^3}3  \left( \frac{\eta_1}{\theta_0 L_0} \right)  + \frac{|\hat\theta^5|}{20} \right]
  + \frac{|\hat\theta^3|}3 \frac{\beta_1}{L_0}. \nonumber
\end{align}
\end{widetext}

\subsection{Degrees of freedom and optimization\label{sec:rb_dof}}

Besides the characteristics of the main bending magnet, the emittance ratio $F$ of the reverse-bend cell in Eq.~\eqref{eq:fullemit} is fully defined by $\beta_0$, $\eta_0$ at the main-bend center, $\beta_1$, $\eta_1$ at the reverse-bend center (all lengths and angles in units of $L_0, \theta_0$), and $\hat\theta$. These five parameters are also sufficient to find the half-cell phase advance $\phi$. One degree of freedom is absorbed by enforcing $\vec P_0$ and $\vec P_1$ to be located on a circle (equal $\mathcal H_\mathrm{BF}$, Fig.~\ref{fig:twobend_phasecircle}), so that four degrees of freedom remain. 

\begin{figure*}[t]
  \includegraphics[scale=0.85]{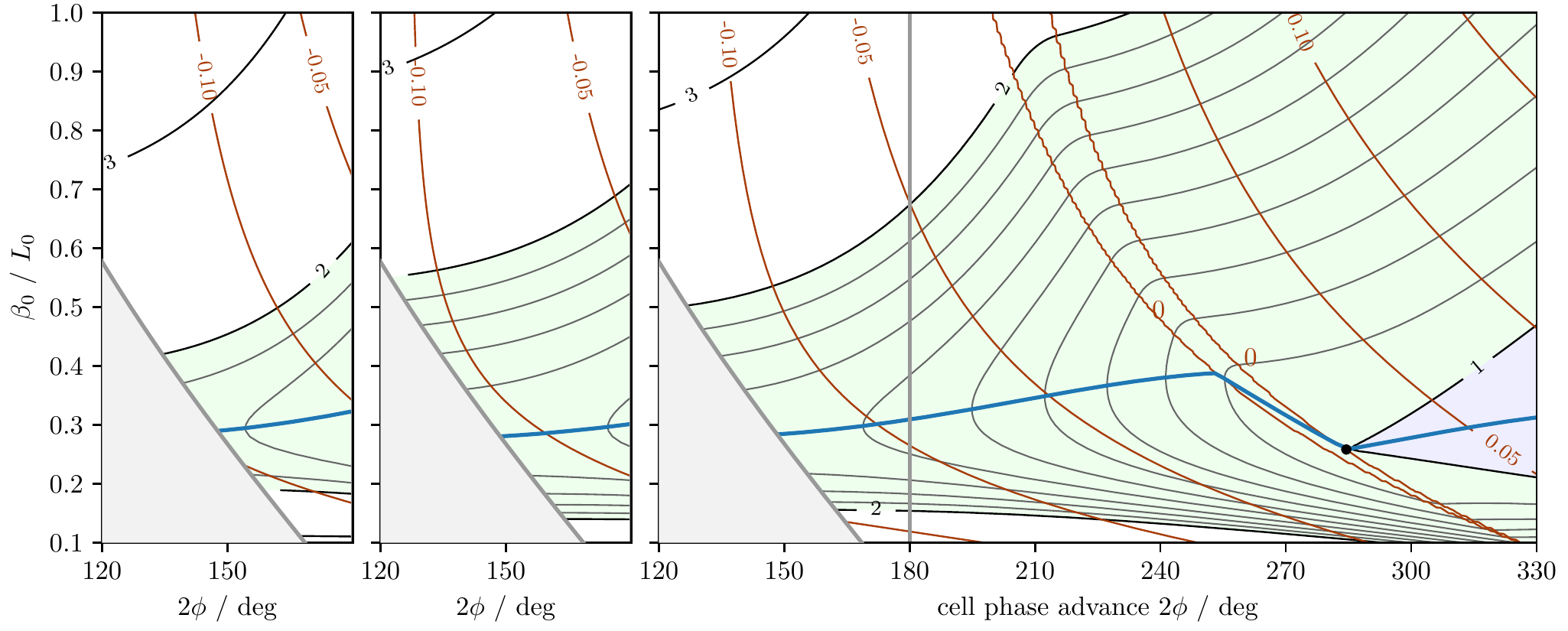}
  \caption{Emittance ratio $F$ for RB cells (see legend in
  Fig.~\ref{fig:IsoMagTme}) with different values of $\beta_1$ at the
  reverse-bend center. From left to right: $\beta_1 = 20 L_0$, $\beta_1 = 40 L_0$, $\beta_1 = 30 L_0$. Red lines show $\theta_1 / \theta_0$ for the respective
  cells in steps of 0.05. At large phase advances, two iso-lines for
  $\theta_1=0$ (actually $\pm 10^{-4}$) delimitate a 'plateau' of vanishing reverse
  bending angle, the black dot at its edge marking the TME condition like in Fig.~\ref{fig:IsoMagTme}.\label{fig:rb_betph}}
\end{figure*}

To obtain some insight into the emittance properties of the cell relative to isomagnetic cells, we choose the free parameters as $(\beta_0, \phi, \beta_1, \hat\theta)$. One may then numerically find the optimal $F$ for a given set of parameters $(\beta_0,\phi,\beta_1)$ with the free parameter $\hat\theta$ using e.g. direct search as performed in the following cases, and show slices of the resulting three-dimensional parameter space.

\subsection{Reverse-bend cells with homogeneous main bend (RB cells)\label{sec:rb}}

For a homogeneous main bend, one can replace $\avg{|\hat b^3| \hat{\mathcal H}}_0$ in
Eq.~\eqref{eq:fullemit} with $\avg{\hat{\mathcal H}}_0$ from Eq.~\eqref{eq:meanH_ctme} and set
  $\avg{\hat b^2}_0 = 1$, obtaining a simplified emittance expression
\begin{align}
  F = \frac{3\sqrt{15}}2 \;\frac{\avg{\hat{\mathcal H}}_0 + \hat\theta^4 \avg{\hat{\mathcal H}}_1}{(1 + \hat\theta)^3 (1 + |\hat\theta|)}. \label{eq:rb_f}
\end{align}
Although $\avg{\hat{\mathcal H}}_1 > \avg{\hat{\mathcal H}}_0$, the $\hat\theta^4 \avg{\hat{\mathcal H}}_1$ summand is small relative to $\avg{\hat{\mathcal H}}_0$.
The major effect on the emittance is then given by $\avg{\hat{\mathcal H}}_0$, as in the TME
cell, and the changes in bending angle described by the denominator.
Thus we can expect that without constraints, the optimal values of $\beta_0, \eta_0$ will only slightly deviate from those of the TME cell.

The results of the numerical optimization for given $\beta_0,\phi_0,\beta_1$ are presented in
Fig.~\ref{fig:rb_betph}. It is visible that two regimes with different sign of
$\theta_1$ exist. For large phase advances, the emittance reduces even
below $F < 1$. This behaviour can be interpreted as
the two-bend cell approximating a double-period TME cell, which would be reached at
$\phi = 2\phi_\mathrm{TME}, \beta_0 = \beta_0^\mathrm{TME}$ and would result in $F = 1/8$.

For sensible phase advances $2\phi < \pi$, the bending angle of the second bend is reversed as expected. Comparing to the considered isomagnetic cells with positive curvature, the $F<2$ region now extends into the range of sensible phase advance, so that lower emittances are feasible.

One can furthermore observe the found relation between $\beta_1$, $\theta_1$, and emittance: low $|\theta_1|$ allows to reach lower emittances, but requires large $\beta_1$ values, while larger absolute reverse-bend angles allow moderate $\beta_1$ values at larger emittances. 

We conclude that at moderate reverse bending angles $|\theta_1| < \theta_0 / 5$, significantly lower emittances than with an LGB or relaxed TME cell are possible for the interesting range of phase advances due to prevention of optical mismatch at the main bend, which is a clear advantage of this cell type.

\subsection{Reverse-bend example cell with homogeneous main bend\label{sec:RbExample}}

\begin{figure}[hb]
  \includegraphics[scale=0.85]{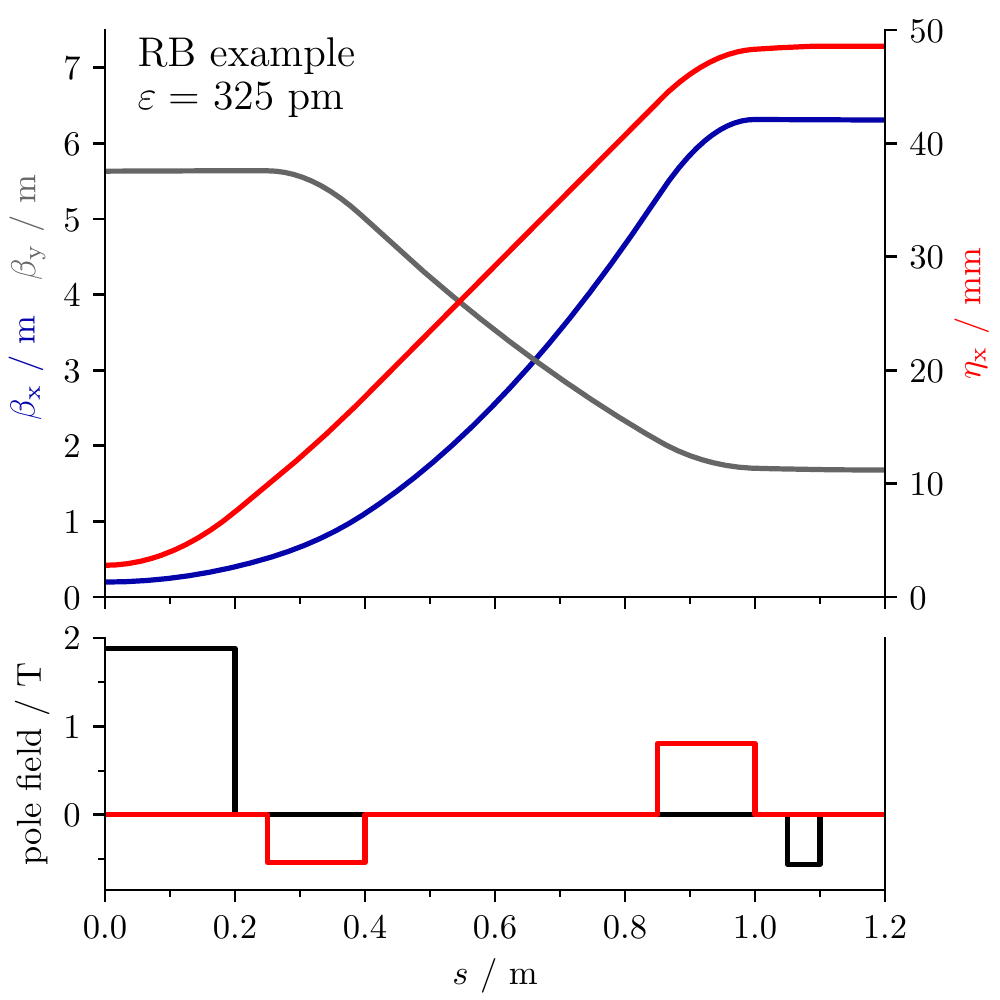}
  \caption{\label{fig:cellc}Example RB cell with homogeneous main bend, utilizing a reverse bend for suppression of dispersion at main bend. (See Fig.~\ref{fig:cella} for legend.)}
\end{figure}

Adding a small reverse bend of length $L_1=\SI{0.05}m$ ($L_1=L_0/4$) to the half-cell from sec.~\ref{sec:lgbExample} increases its length, but effectively reduces the dispersion at the main bend center, see
Fig.~\ref{fig:cellc}. Note that the reverse bend is shifted from the half-cell end, allowing for the installation of a sextupole magnet at that symmetry point and leading to half-cell length of \SI{1.2}m.

Here the RB angle was set to $\theta_1=-0.2^\circ$, and the angle of the main
bend $\theta_0$ was increased by the same amount in order to maintain the total
cell deflection ($\hat\theta=0.074$). As a consequence the field of the main
bend is higher since its length was maintained. At unchanged phase advances
$\phi,\phi_\mathrm y$ (resulting in increased $\beta_0=\SI{0.198}m$), the emittance shrinks to
$\varepsilon =\SI{325}{pm}$ (or $F=2.69$), which is 72~\%\ of the value from the relaxed TME cell example (sec.~\ref{sec:tmeExample}).

\subsection{Reverse-bend cells with LGB as main bend (LGB/RB cell)\label{sec:lgbrb}}

Lowering $\eta_0$ by means of RBs eventually enables an LGB as main bend to efficiently reduce the emittance: a central peak of high field concentrates most bending and quantum emission in a region of small ${\cal H}$, while decay of field strength towards the edges compensates for the inevitable growth of ${\cal H}$, thus minimizing the radiation integral $I_5$. 
So, the final model to be investigated is a unit cell composed from LGB and RB.

The necessary computations have already been carried out in sec.~\ref{sec:lgb}, resulting in Eq.~\eqref{eq:lgb_b3H}, Eq.~\eqref{eq:lgb_etavee}, and in the beginning of this section, resulting in Eq.~\eqref{eq:fullemit}. We again use the inverse-distance scaling magnet (IDM) with field enhancement factor $R=2$ as an example and insert its parameters into the numerical optimization procedure for RB cells.

\begin{figure*}[t]
  \includegraphics[scale=0.85]{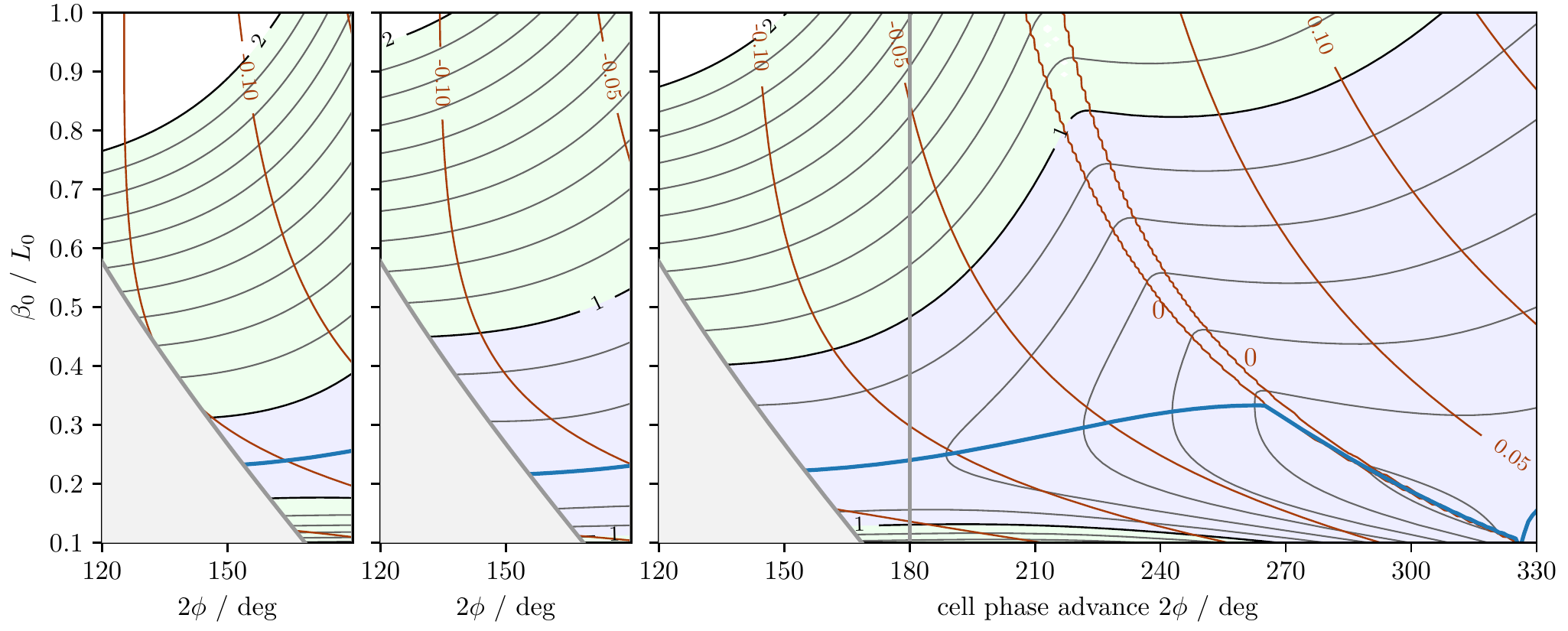}
  \caption{Emittance ratio $F$ for LGB/RB cells with IDM magnet at $R=2$ (see
    legend in Fig.~\ref{fig:rb_betph}) with different values of $\beta_1$ at the
    reverse-bend center. From left to right: $\beta_1 = 20 L_0$, $\beta_1 = 40 L_0$, $\beta_1 = 30 L_0$.\label{fig:RbCellsIdmF}}
\end{figure*}

The results of this procedure are shown in Fig.~\ref{fig:RbCellsIdmF}. Like for RB cells, the region of low emittances is extended into the range of phase advances $2\phi < \pi$ by virtue of the reverse-bend scheme. This 'mapping' of optimal main-bend parameters to lower phase advances is also beneficial for exploitation of the emittance reduction enabled by using LGB.

As can be observed in comparison with Fig.~\ref{fig:rb_betph}, the emittance of an LGB/RB cell can be significantly smaller than that of a RB cell and potentially even reach $F\leq 1$. This is similar to the case of an LGB in an isomagnetic cell without constraints (sec.~\ref{sec:optUnbounded}), but the reverse bend in an LGB cell is more effective, allowing to reach similar emittances with a technically feasible field enhancement factor of only $R=2$.

We conclude that the figures of merit being low cell phase advance, optics matching in the main bend and exploitation of LGB characteristics cannot not be reached simultaneously using relaxed TME, LGB ($b \geq 0$), and RB cells, and that the LGB/RB cell type yields superior performance to the other investigated cell types. To apply further boundary conditions on the cell design, we need to assume specific properties on the quadrupole array in the cell interior (see appendix \ref{app:bendFreeMatch}).

\subsection{Reverse-bend example cell with LGB as main bend\label{sec:LgbRbExample}}

\begin{figure}[b]
  \includegraphics[scale=0.85]{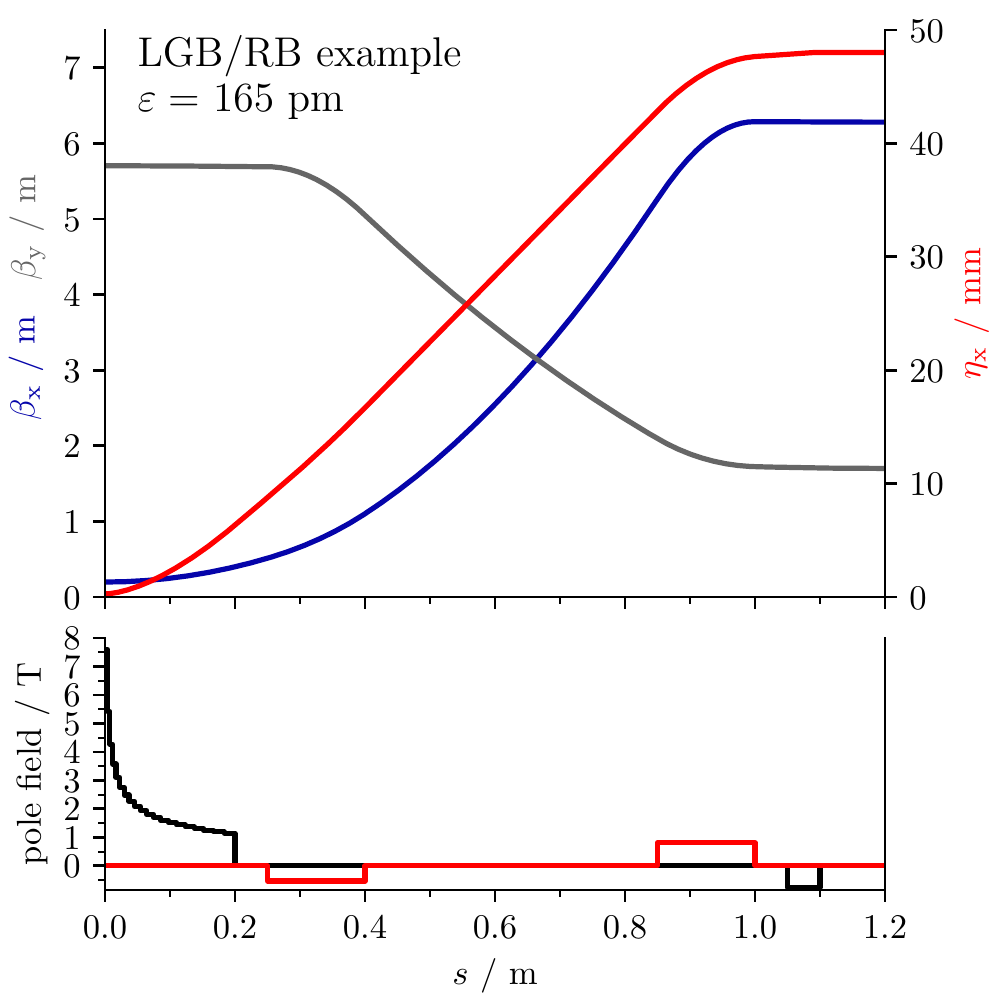}
  \includegraphics[scale=0.85]{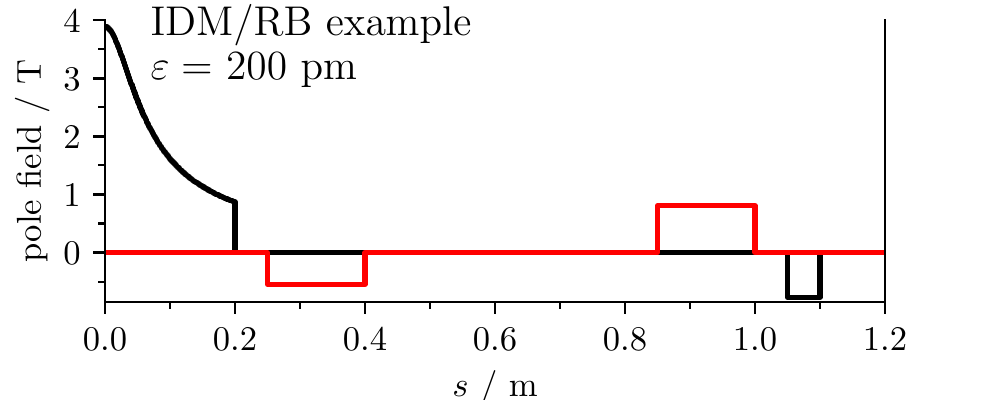}
  \caption{\label{fig:celld} Top: Field variation in a longitudinal gradient bend efficiently exploits the suppression of dispersion enabled by the reverse bend. (See Fig.~\ref{fig:cella} for legend.)
  \label{fig:celldalt} Bottom: Magnetic field of an LGB/RB example cell utilizing an IDM with $R=2$.}
\end{figure}

With an RB providing the means to suppress the dispersion (and with it the invariant $\cal H$) at the main bend center, optimization of the field variation now efficiently suppresses the fifth radiation integral by pushing the central field peak to very high values (see Fig.~\ref{fig:celld}). LGB profile and RB angle were optimized in common, since a high field peak calls for central dispersion close to zero, resulting in $\theta_1=-0.275^\circ$. At unchanged tunes $\phi,\phi_\mathrm y$, the emittance is reduced to \SI{165}{pm} (or $F=1.36$), which is only 36~\%\ of the relaxed TME cell example (sec.~\ref{sec:tmeExample}). However, the narrow high field peak does not correspond to a realistic magnet design.

This problem is partially be circumvented by using an IDM magnet as LGB (IDM/RB example, Fig.~\ref{fig:celldalt}). Although the emittance $\varepsilon = \SI{200}{pm}$ (or $F=1.65$) of this cell is still relatively small, a moderate field enhancement factor $R=2$ corresponds to a peak field $B_\mathrm{max} \approx \SI 4T$ in our example cell.

As $B_\mathrm{max} \propto 1/L_0$ for fixed field enhancement, the main bend length would need to be doubled to obtain similar emittances using a normal-conducting magnet, also beneficially reducing $\beta_0 < L_0 / 2$. To maintain the focusing constraints of the considered example cell, the main bend would be required to spatially overlap with the vertically focusing quadrupole.

\section{The SLS 2.0 unit cell\label{sec:realcells}}

In the previous sections, unit cells made of separate-function magnets have been considered, optimizing the emittance using the radiation integrals $I_5$ (which only affects transverse emittance) and $I_2$.

In this section, we lift the constraint of using separate-function magnets, which was required for our simplified dispersion matching and emittance model. This opens up the possibility of manipulating the emittance
via the horizontal damping partition $J_\mathrm x=1-I_4/I_2$, as is explained in more detail in \cite{antibend}. The emittance is proportional to
\begin{align}
  F &\propto \frac{I_5}{I_2 J_\mathrm x} = \frac{I_5}{I_2 - I_4} \quad \text{with}
  \quad I_4 = \int \eta b (b^2 + 2 k) \totd s,
\end{align}
where $k = \totd b / \totd x$ is the normalized transverse gradient, $k > 0$ resulting in horizontal focusing.

It is thus possible to decrease $F$ by introducing simultaneous bending and focusing so that $\eta b\cdot k < 0$. For positive dispersion, this requires vertical focusing in the main bend and horizontal focusing in the reverse bend; conveniently the gradients perform the function of a quadrupole doublet as required for focusing at sensible phase advances $2\phi < \pi$.

In addition, the combination of the LGB with a quadrupole effectively increases $L_0$ and thus allows high field enhancement at lower absolute curvatures (see sec.~\ref{sec:LgbRbExample}). The transverse gradient is included near the magnet ends with lower curvatures, where it is technically feasible. 

\begin{figure}[htb]
  \includegraphics[scale=0.85]{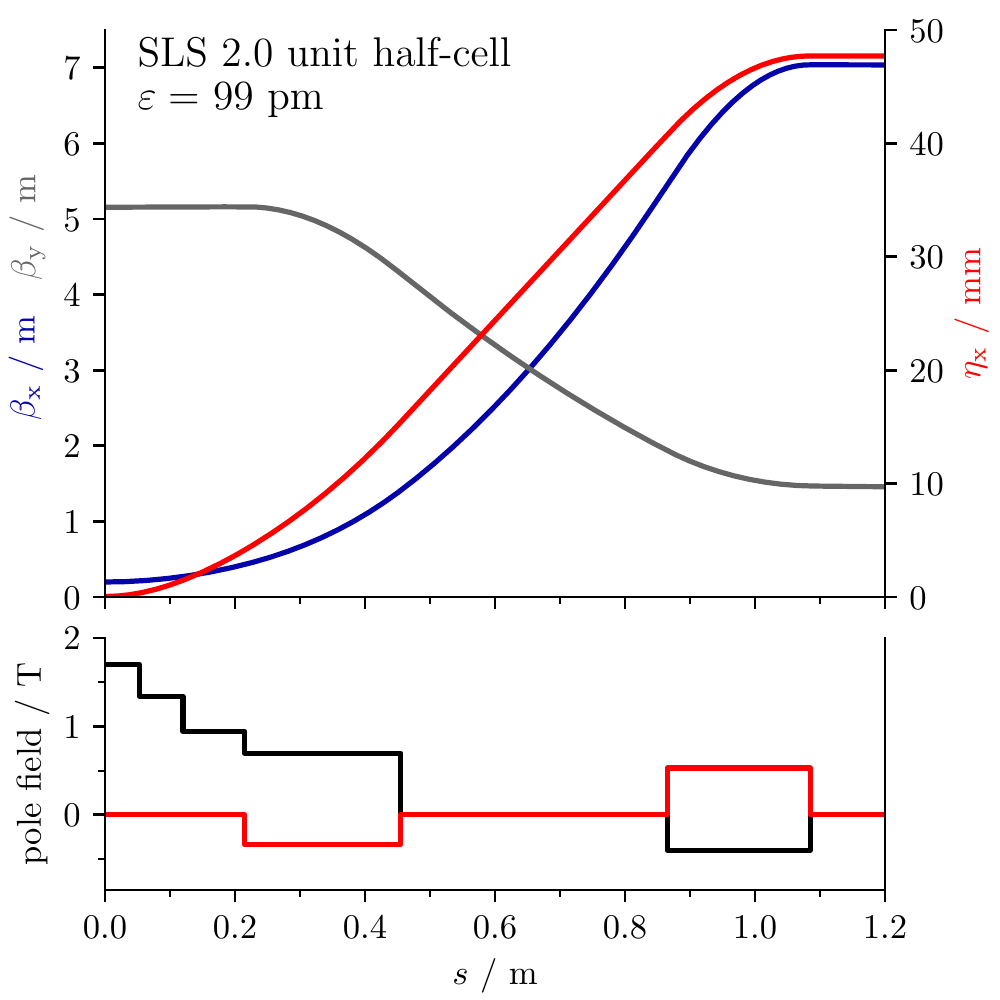}
  \caption{\label{fig:cells}The unit cell of the SLS upgrade lattice ``SLS 2.0'' represents a fully optimized, real LGB/RB cell. (See Fig.~\ref{fig:cella} for legend.)}
\end{figure}

Eventually, a real lattice cell based on the aforementioned combined-function magnets is shown in Fig.~\ref{fig:cells}: it is the unit cell of the new storage ring for the upgrade of the Swiss Light Source SLS, named ``SLS~2.0''~\cite{sls2-jsr,sls2-cdr,sls2-ipac18}. Here the tunes were slightly shifted to $Q_x=0.4285\approx 3/7$ and $Q_y=0.1429\approx 1/7$ for optimal cancellation of sextupole and octupole resonances over an arc made from seven cells~\cite{johan-sls2rep}. The RB is merged with the horizontally focusing quadrupole~-- essentially the RB is a quadrupole shifted radially away from the storage ring center. The LGB is a permanent magnet with moderate peak field. The low field in the end pieces provides margin to introduce a transverse gradient for vertical focusing. The bending angles are $\theta_0=1.93^\circ+1.2^\circ$ (central half LGB $+$ end piece) and $\theta_1=-0.63^\circ$ (RB), giving $5^\circ$ deflection for the complete unit cell.

Since time of flight effects in main and reverse bends compensate to some extent, LGB/RB cells become almost isochronous when tuned to the minimum emittance for given phase advance, which is an attractive feature for the use in damping rings. In light sources, however, longer bunches are required to provide adequate beam lifetime and instability thresholds. Therefore the RB angle in the SLS~2.0 cell is increased beyond its optimum value in order to realize a sufficient negative momentum compaction factor while accepting the emittance to be 9\%\ larger than its possible minimum value.

\begin{figure}[htb]
\includegraphics[scale=0.85]{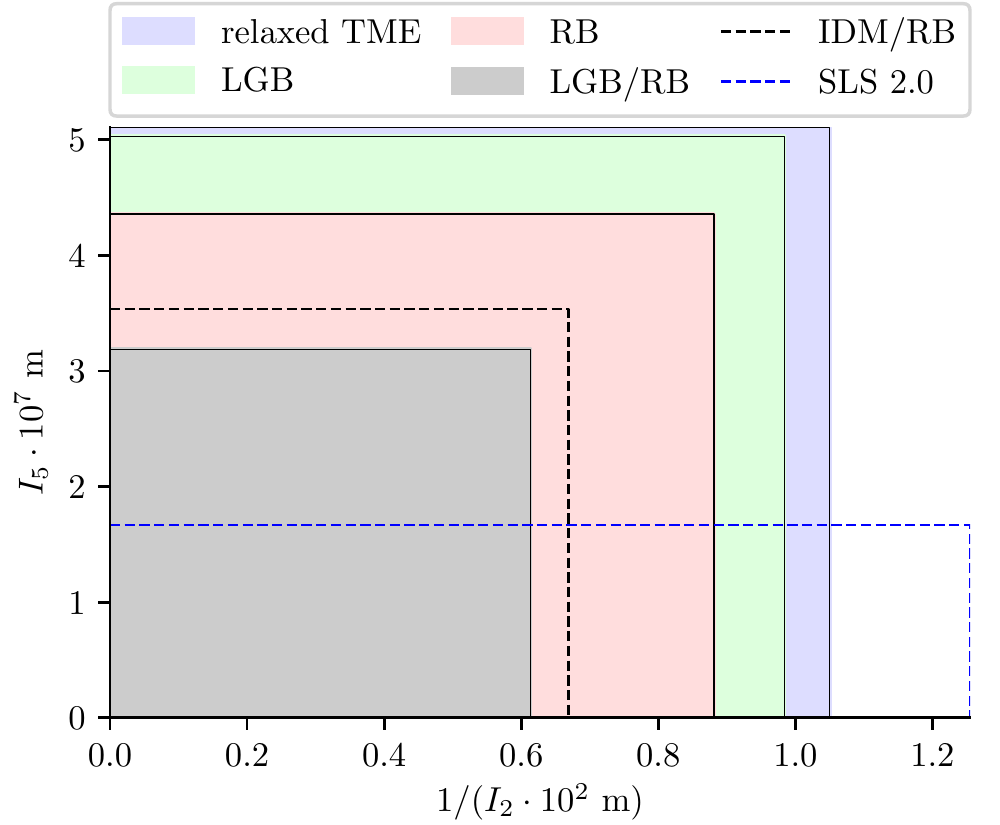}
\caption{\label{fig:histo}Comparison of radiation integrals $I_5$ and $I_2$ for the example half-cells including SLS 2.0. The area of the respective rectangles is proportional to the emittance when approximating $J_\mathrm x=1$.}
\end{figure}

A comparison of $I_2$ and $I_5$ radiation integrals for all considered
example cells is shown in Fig.~\ref{fig:histo}. The fraction of radiation integrals $I_5/I_2$ of the SLS 2.0
unit cell is located between that of the LGB/RB and IDM/RB examples. Both
radiation integrals are lower in the SLS 2.0 case, since the fields on average are lower due to the increased effective length $L_0$ of the main bend. Thus, in comparison to the example cells, the major emittance reduction in terms of $I_5/I_2$ is due to reduction of $I_5$, which is beneficial as this integral only affects transverse emittances.

The final emittance of this unit cell is further reduced through the increased damping partition number $J_\mathrm x$.
A positive transverse gradient in the RB and a negative gradient in the main bend both shift the damping partition in favor of the horizontal dimension resulting in $J_\mathrm x =1.795$, whereas $|J_\mathrm x-1| < \num{2e-3}$ in all previous example cases (see drift-space assumption, sec.~\ref{sec:IsoMag}).

After application of all described optimizations and design choices, a unit cell emittance of $\varepsilon = \SI{99}{pm}$ is obtained for a beam energy of \SI{2.4}{GeV}, which corresponds to a TME emittance ratio $F=0.82$.

\section{Conclusion}

\begin{enumerate}
  \item We reviewed the commonly known fact that the TME condition is not realized in MBA unit cells due its high phase advance $2 \phi > \pi$. Instead, relaxed TME cells with sensible phase advances $2 \phi < \pi$ are used. The resulting optical mismatch in the bend causes an emittance increase of such cells ($F > 2.45$).
  \item We demonstrated that LGB shapes \cite{gradientbend} exist which provide significant emittance reduction at their optimal, very high phase advances $\phi > \phi_\mathrm{TME}$, but which actually provide only marginal emittance reduction compared to a a relaxed TME cell for sensible phase advances. Even without specification of any particular shape, there seems to exist a principal limit $F > 2$ of achievable emittance reduction by LGBs for sensible phase advances.
  \item Free-form optimization of LGB curvatures without constraints on curvature polarity results in two regimes: positive curvatures in the magnet center and negative curvature at its ends, i.e.~the bend naturally splits up into a main bend and a reverse bend.
  \item The concept of the reverse-bend cell \cite{antibend} has been revisited. It has been shown that for sensible phase advances, the reverse-bend cell is able to provide significantly lower emittances than relaxed TME cells through suppression of dispersion in the main bend center ($F < 2$ is possible for $2\phi<\pi$).
  \item When combining a RB with an LGB (LGB/RB cell), its potential in emittance reduction can be exploited and even lower emittance values $F < 1$ are also possible for sensible phase advances $2\phi < \pi$:
a central peak of high field concentrates most bending in the region of suppressed dispersion, while decay of field strength towards the edges compensates for the inevitable growth of dispersion, thus minimizing the quantum excitation integral. 
\end{enumerate}

\begin{acknowledgments}
  The authors thank M. Aiba for thorough reading and commenting on the
  manuscript, as well as J. Chrin and M. B\"{o}ge for useful discussions.
\end{acknowledgments}

\appendix
\section{Emittance coefficients for magnets\label{app:eint}}

With the drift-space assumption $\beta(s) = \beta_0 + s^2 / \beta_0$ and Eq.~\eqref{eq:etadef}, and
omitting the indices of $L_q,\theta_q$ for convenience, the
normalized dispersion invariant Eq.~\eqref{eq:tmelike}, Eq.~\eqref{eq:normH} evaluates to (see \cite{gradientbend})
\begin{align}
  \hat{\mathcal H} &= \frac H{L \theta^2} \label{eq:A1}\\
  &= \frac L{\beta_q} \left[ \left( \frac{\eta_q}{\theta L} \right)^2 - 2 \frac{\eta_q}{\theta L} v(s) + v^2(s) \right] + \frac{\beta_q}L \left(\frac{\eta^\prime(s)}{\theta}\right)^2. \nonumber 
\end{align}
where we introduced the lever function
\begin{align}
  \label{eq:normLever}
  v(s) = \frac 1{\theta L} \left( s \eta^\prime(s) - \int\limits_0^s \eta^\prime(\tilde s) \totd \tilde s \right)
  = \frac 1{\theta L} \int\limits_0^s \tilde s \; b(\tilde s) \totd \tilde s .
\end{align}

Multiplication of Eq.~\eqref{eq:A1} and averaging results in $\avg{|\hat b^3| \hat{\mathcal H}}$. The prefactor of $\beta_q/L$ modifies to
\begin{align}
   C &= \avg{|\hat b^3| \left( \frac{\eta'}{\theta} \right)^2},
\end{align}
while the prefactor of $L/\beta_q$ modifies to
\begin{align}
  \avg{|\hat b^3|}\left(\frac{\eta_0}{\theta L}\right)^2 - 2 \avg{|\hat b^3|v}\left(\frac{\eta_0}{\theta L}\right) + \avg{|\hat b^3|v^2}.
\end{align}
These remaining four averages in $\avg{|\hat b^3| \hat{\mathcal H}}$ are only dependent on $\hat b(s)$.

To furthermore compute the phase advance for any magnet, a relation between
$\eta_\vee$ and $\eta_0$ (or $\eta_\wedge$ and $\eta_1$) is established by the thin-dipole dispersion difference
\begin{align}
  \eta_0 &= \eta_\vee + V \theta_0 L_0 & \text{with } V &= v(L_0).
\end{align}

For magnets with positive orbit curvature, $b, \; \eta', \; v$ are all positive; in this case, 
the lever function and its final value are related as $V = v(L_0) = \max v(s)$.
For Eq.~\eqref{eq:lgb_b3Hphi} positive curvature thus implies
\begin{align}
  \tilde A &= 2 \left( \avg{|\hat b^3|} V - \avg{|\hat b^3|v} \right) = 2 \avg{|\hat b^3| (V-v)} > 0. 
\end{align}

\subsection{Homogeneous magnet \label{app:hom}}

For this magnet type, extending from $0 < s < L$,
\begin{align}
  \hat b(s) &= 1, &
  \frac{\eta'(s)}{\theta} &= \frac sL, &
  v(s) &= \frac 12 \left( \frac sL \right)^2.
\end{align}
Note that the sign of bending angle is normalized out of $\hat b$, so that it can also be used for reverse bends. This results in
\begin{align}
  \avg{|\hat b^3|} &= 1, & 
  \avg{|\hat b^3|v} &= \frac 16, &
  \avg{|\hat b^3|v^2} &= \frac 1{20}, \\
  \avg{\hat b^2} &= 1, & 
  V &= \frac 12, &
  C &= \frac 13.\nonumber
\end{align}

\subsection{Inverse-distance scaling magnet (IDM)}

Using the inverse distance $w(s)=\sqrt{1+(s/h)^2}$, the dispersion-related functions of the IDM are obtained as
\begin{align}
  \frac{\eta'(s)}{\theta} &= \frac{\arsinh(s/h)}{\arsinh(L/h)}, &
  v(s) = R&(h/L)^2 \left( w(s) - 1 \right), \nonumber\\
  V &= R (h/L)^2 \left(W - 1\right) & \text{with } W &= w(L).
\end{align}

Most emittance coefficients can be expressed via averages of the form
$\avg{w^{-m}}$ with positive integer $m$,
\begin{align}
  \avg{\hat b^2} &= R^2 \avg{w^{-2}}, \quad \avg{|\hat b^3|} = R^3 \avg{w^{-3}}, \nonumber\\
  \avg{|\hat b^3|v} &= R^4 (h/L)^2 \left( \avg{w^{-2}} - \avg{w^{-3}} \right), \\
  \avg{|\hat b^3|v^2} &= R^5 (h/L)^4 \left( \avg{w^{-1}} - 2\avg{w^{-2}} + \avg{w^{-3}}\right).\nonumber
\end{align}
The case $m=1$ of this average evaluates to $\avg{w^{-1}} = \arsinh(L/h) \cdot
h/L$. For larger $m$, one may use the substitution $u=\arctan(s/h)$ so that the integrand transforms to $\cos^{m-2} u$; the integral is then solved by recursive application of the cosine reduction formula.

The remaining coefficient for the emittance is
\begin{align}
  C = \frac{R^3}{\arsinh^2(L/h)} \avg{\frac{\arsinh^2(s/h)}{w^3}}.
\end{align}
Introducing the abbreviations $a = \arsinh(L/h)$,
the average in $C$ evaluates to
\begin{widetext}
\begin{align}
  \frac hL \left( \dilog\left( 1+\mathrm e^{-2a} \right) + a^2 \frac{L/h - W}W - 2a \ln\left( 1+\mathrm e^{-2a} \right) + \frac{\pi^2}{12} \right) \\
  \text{with the dilogarithm} \quad \dilog z = \int_0^z \frac{\ln t}{1-t} \totd t.\nonumber
\end{align}
\end{widetext}

\subsection{Free-form LGB shapes\label{app:freeform}}

For numerical integration, we split $b(s)$ into a piece-wise constant function
with $Q$ pieces
\begin{align}
  \hat b(s) = \underline b_q \text{ for } s_{q-1} < s < s_q,
\end{align}
with $s_0=0, s_Q = L,$ and the values $s_q$ being equidistant with
$\Delta s = L/Q$. In that case, $\eta^\prime(s)$ is a piece-wise linear function with
\begin{align}
  \frac{\eta^\prime(s_p)}{\theta} &= \frac 1Q \sum_{q=1}^p \underline b_q
\end{align}
being a cumulative sum ($0 < p \leq Q$). The lever function Eq.~\eqref{eq:normLever} follows as
\begin{align}
  v(s_p) &= \frac 1Q \sum_{q=1}^p \frac{s_q + s_{q-1}}{2 L} \; \underline b_q.
\end{align}
The emittance coefficients have more complicated dependencies on the $\underline b_q$ values, and one may compute them using the numerical average in good approximation, subsequently calculating $F$ using Eq.~\eqref{eq:lgb_b3Hphi}.

Note that all described calculations (also including the normalization requirement $\eta^\prime(L) = \theta$) are differentiable transformations of the parameters $\underline b_q$ , which allows to compute the gradient of the discretized functional $\totd F / \totd \underline b_q$ for all $q$.

\section{Matching the bend-free region for given parameters\label{app:bendFreeMatch}}

Of the four free parameters of a reverse-bend cell $\phi,\beta_0,\beta_1,\theta_1/\theta_0$, the former three parameters define the half-cell transfer matrix $\mathbf T$ via Eq.~\eqref{eq:T_half}, and the ratio $\theta_1/\theta_0$ defines the magnet lengths $L_0,L_1$ via Eq.~\eqref{eq:L0L1}. The magnet transfer matrices are assumed as drift spaces $\mathbf D_0, \mathbf D_1$. In consequence, the horizontal transfer matrix of the bend-free region can be calculated as
\begin{align}
  \label{eq:matchM} \mathbf M &= \mathbf D_1^ {-1} \mathbf B_1 \mathbf R(\phi) \mathbf B_0^{-1} \mathbf D_0^{-1}\\
  &= \frac 1{\sqrt{\beta_0 \beta_1}}
  \begin{pmatrix}
    \beta_1 & -L_1 \\ 0 & 1 
  \end{pmatrix} \mathbf R(\phi)
  \begin{pmatrix}
    1 & -L_0 \\ 0 & \beta_0
  \end{pmatrix}. \nonumber  
\end{align}
To obtain a half-cell for given $\phi, \beta_0, \beta_1$, the transfer matrix $\mathbf M$ of the bend-free region must thus be matched by a given array of quadrupoles, while also guaranteeing vertical stability. This naturally requires two or more quadrupoles, separated by drift spaces.

The conceptually simplest example is a Galilean telescope, modeled using two thin lenses with focal lengths
of different signs $f_0, f_1$ directly attached to the ends of the bending
magnets, and an intermediate drift space of length $d$. As the transfer matrix
must be $\mathbf M$, one obtains
\begin{align}
  d &= M_{12}, & \frac d{f_0} &= 1 - M_{11}, & \frac d{f_1} &= 1 - M_{22}.
\end{align}
The sign of $f_0$ is given by calculation of $M_{11}$ from Eq.~\eqref{eq:matchM}
\begin{align}
  M_{11} = \sqrt{\frac{\beta_1}{\beta_0}} \cos \phi + \frac{L_1}{\sqrt{\beta_0\beta_1}} \sin \phi.
\end{align}
For the interesting range of phase advances $\phi < \pi / 2$ and under the assumption $\beta_1 / \beta_0 \gg 1$, we conclude that the focal length of the main-bend lens $f_0$ is negative.

\end{document}